\documentclass[a4paper,11pt]{article}
\pdfoutput=1 

\usepackage{jheppub} 
\everymath{\displaystyle}

\usepackage[T1]{fontenc} 
\usepackage{braket}
\usepackage{upgreek}
\usepackage{subfigure}
\usepackage{xcolor}
\usepackage{bm}
\usepackage{multirow}
\usepackage[english]{babel}
 
\usepackage{amsmath,amsfonts}
\usepackage{amssymb}
\usepackage{subcaption}
\usepackage{graphicx}
\usepackage{parskip}
 
\newcommand{\drawsquare}[2]{\hbox{%
\rule{#2pt}{#1pt}\hskip-#2pt
\rule{#1pt}{#2pt}\hskip-#1pt
\rule[#1pt]{#1pt}{#2pt}}\rule[#1pt]{#2pt}{#2pt}\hskip-#2pt
\rule{#2pt}{#1pt}}
\newcommand{\Yfund}{\raisebox{-.5pt}{\drawsquare{6.5}{0.4}}}
\newcommand{\Ysymm}{\Yfund\hskip-0.4pt%
                    \Yfund}
\def\symm{\Ysymm}
\def\bsymm{\overline{\Ysymm}}

\def\drawbox#1#2{\hrule height#2pt
        \hbox{\vrule width#2pt height#1pt \kern#1pt
              \vrule width#2pt}
              \hrule height#2pt}

\def\Asym#1#2{\vcenter{\vbox{\drawbox{#1}{#2}
              \kern-#2pt       
              \drawbox{#1}{#2}}}}

\def\asymm{\Asym{6.4}{0.3}}

\def\basymm{\overline{\asymm}}

\def\Acknowledgements{\bigskip  \bigskip {\begin{center}
              \bf Acknowledgments \end{center}}}

\newcommand {\beq} {\begin{equation}}
\newcommand {\eeq} {\end{equation}}
 \newcommand{\be}{\begin{eqnarray}}
\newcommand{\ee}{\end{eqnarray}}

\usepackage{orcidlink}

\title{  \centering   { \Huge Two-index SU(N) theories }
\\~\\    QCD, Orientifolds, Super Yang-Mills, Lattice  
\\   and  
\\  Steven Weinberg's   $\pi \pi$ scattering legacy}

\author{Francesco Sannino
\emailAdd{sannino@qtc.sdu.dk}
\affiliation{\small Quantum  Theory Center ($\hbar$QTC) \& D-IAS, Southern Denmark Univ., Campusvej 55, 5230 Odense M, Denmark}
\affiliation{\small Scuola Superiore Meridionale, Largo S. Marcellino, 10, 80138 Napoli NA, Italy}
\affiliation{\small Dept. of Physics E. Pancini, Università di Napoli Federico II, via Cintia, 80126 Napoli, Italy}
\affiliation{\small INFN sezione di Napoli, via Cintia, 80126 Napoli, Italy}

\abstract{I review and improve on how two-index SU(N) gauge-fermion theories help access salient information about the large $N$ vacuum and spectrum of QCD, super Yang Mills and meson-meson scattering. The interplay with recent lattice simulations will be employed to deduce the size of $1/N^2$ corrections. Through the meson-meson scattering   analysis I will honor Steven Weinberg's memory by showing how two-index   extrapolations naturally accommodate the appearance of tetraquarks states crucial to unitarize meson-meson scattering at low energies.   }

\makeatletter
\gdef\@fpheader{}
\makeatother

\begin{document}

\maketitle

\newpage 

\section{Prologue}
\label{uno}

Steven Weinberg, a titan of physics, profoundly impacted our understanding of the fundamental forces of Nature. His monumental contribution spans from particle physics to cosmology. Weinberg's work not only advanced the frontiers of theoretical physics but also inspired generations of scientists. His legacy will continue to influence our understanding of the building blocks of nature and the scientific community at large for the time to come. It is hard not to be both amazed and intimidated by Weinberg's work, always written with such a rigor and clarity but at the same time pragmatic and to the point.  Weinberg's defining work \cite{Weinberg:1967tq} alongside Glashow and Salam, is the discovery of the electroweak theory, a model blessed by Nature \cite{Aad:2012tfa}. My past and present research work, like the one of several colleagues, has drawn inspiration by many of Weinberg's papers from his pioneering work on spontaneous symmetry breaking \cite{Goldstone:1961eq} to $\pi \pi$ scattering \cite{Weinberg:1966kf},  effective lagrangians \cite{Weinberg:1978kz},  dynamical \cite{Weinberg:1975gm} and radiative \cite{Gildener:1976ih} symmetry breaking \cite{Weinberg:1975gm}, high temperature symmetry non restoration \cite{Weinberg:1974hy}, etc. Rather than even attempting the impossible, for me, task to provide a comprehensive review of Weinberg's immense body of work I will discuss one of his late research papers about "Tetraquarks mesons in large-$N$ Quantum Chromodynamics"  \cite{Weinberg:2013cfa} and show how  to address his main conclusions efficiently and elegantly using an alternative large number of colors limit. I will also review and improve upon the use of the alternative framework to determine spectral properties of QCD, and more generally of $SU(N)$ gauge theories with two-index matter, of immediate relevance for present and future  lattice simulations and model building. 

Despite the work of physicists of the caliber of Weinberg,  strongly coupled dynamics continues to challenge our  understanding of nature. It is for this reason that several methodologies have been devised to tackle different corners of strongly coupled quantum field theories from effective approaches \cite{Weinberg:1978kz,Gasser:1984gg,Gasser:1987ah} to large number of colors expansions \cite{tHooft:1973alw,Witten:1980sp,Corrigan:1979xf,Coleman_1985}. These methodologies are often fused to increase their impact. A renown example is the 't Hooft ~\cite{tHooft:1973alw,Witten:1980sp} large number of colors limit. Here one can  explain several properties of QCD with several caveats. For, example, since quark loops are suppressed at large $N$ the properties of
the $\eta'$-meson are not properly captured. The latter acquires mass via the quantum axial anomaly which is suppressed in 't Hooft large $N$ limit.  Additionally,  Coleman in his Erice lectures \cite{Coleman_1985} argued that QCD, in the 't Hooft extrapolation, does not feature tetraquark mesons, i.e. exotic mesons made by a pair of quarks and antiquarks.  However, these states do play a crucial role in unitarizing  pion pion scattering as I had shown during my PhD work  \cite{Sannino:1995ik,Harada:1995dc} and, furthermore, Jaffe \cite{Jaffe:1976ig} had already introduced them  to group-theoretically accommodate the low energy scalar spectrum. Tetraquark states have been  discussed later in \cite{Close:2002zu,Braaten:2003he,Pelaez:2003dy,Close:2003sg,Maiani:2004uc,Bignamini:2009sk,Ali:2011ug,Achasov:2012kk}.  

\begin{figure}[h!]
    \centering
    \includegraphics[width=.75\linewidth]{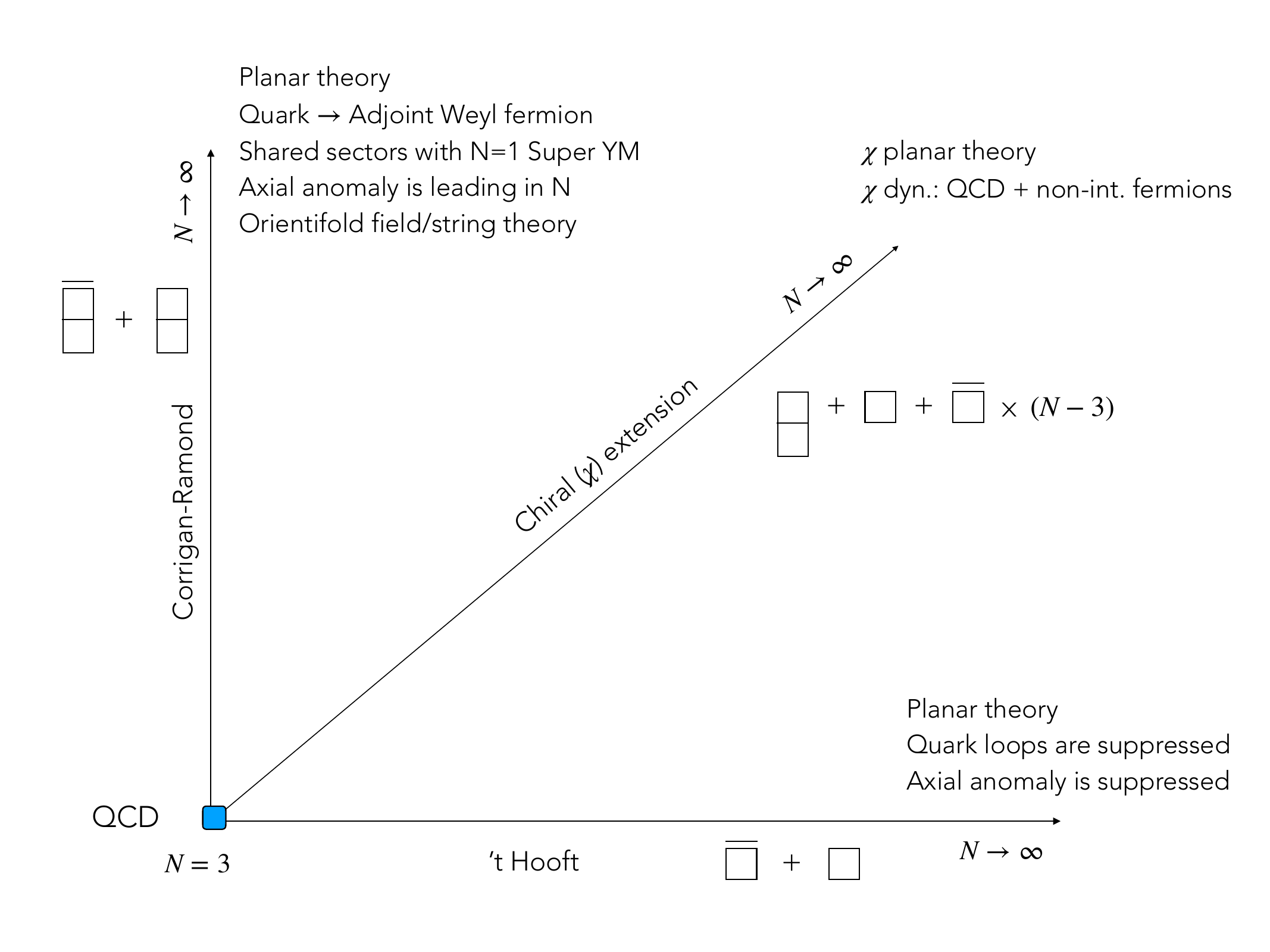}
    \caption{Diagram summarizing the three large $N$ QCD extensions, their matter contents and salient large $N$ properties. This is the improved version of the similar figure first appeared in our earlier work \cite{Ryttov:2005na}.  
}
    \label{diagram-figure}  
\end{figure}

The alternative large $N$ limit was pioneered by Corrigan and Ramond (CR) \cite{Corrigan:1979xf} soon after 't Hooft seminal work, introducing quarks transforming according to 
 the two-index antisymmetric representation of the gauge
group. By construction the two type of QCD extensions  coincide for $N=3$ but  differ at large  $N$. The CR expansion was introduced to amend the 't Hooft overzealous suppression of quark loops as $N$ approaches infinity. The  CR large $N$ rules and physical applications for scattering amplitudes were investigated in \cite{Kiritsis:1989ge,Sannino:2007yp}. As I  shall show later these rules elucidate a number of physical properties of the QCD spectrum  such as the nature of the lowest lying composite scalar known as $f_0(500)$ as a tetraquark state \cite{Sannino:2007yp}.  Besides the 't Hooft and CR vector-like extrapolations of QCD, a chiral extension was discovered in \cite{Ryttov:2005na}. In the chiral extension one splits the two Weyl fermions constituting a QCD quark and assigns one Weyl to the fundamental representation and the other to the two-index antisymmetric representation of the gauge group. This theory is chiral for any $N$ larger than three and maps into the generalized Georgi-Glashow model \cite{PhysRevLett.32.438}  with one vector like fermion. Differently from the 't Hooft and the CR limits in which the idea is to  capture different aspects of QCD, the motivation for a chiral extension is to learn more about the non-perturbative dynamics of the chiral theory itself. In fact, to date lattice simulations of chiral theories are still  challenging \cite{Kaplan:2023pxd,Kaplan:2023pvd}. In \cite{Ryttov:2005na} we learnt that there is a non-perturbative phase of the chiral extension where the dynamical formation of a bilinear condensate  breaks the gauge symmetry  spontaneously and Higgses the theory to ordinary massless one-flavor QCD. The remaining infrared chiral matter is massless and sterile with respect to QCD. The example hints at the possibility that the non-trivial dynamics for chiral gauge theories is vector-like while the residual infrared chiral dynamics is trivial. The distinct large $N$ extensions of QCD are summarised in figure~\ref{diagram-figure} including their salient large $N$ properties. 

Focusing on the CR limit, but simultaneously considering also two-index symmetric theories it was later shown by Armoni, Shifman and Veneziano 
Refs.~\cite{Armoni:2003gp,Armoni:2003fb} that there is a planar  connection among the mesonic
sectors of these theories and  $\mathcal{N}=1$ super
Yang-Mills theory. The two-index theories are also known in string-theory   as {\it orientifold  theories} since they were shown to live on a brane configuration of type 0A string theory
\cite{Armoni:1999gc,Angelantonj:1999qg}  consisting of NS5
branes, D4 branes and an {\it orientifold} plane. It is useful to summarize the perturbative spectrum of the two-orientifold theories in the table~\ref{table}.  
\begin{table}[b]
\begin{center}
\begin{minipage}{2.2in}
\begin{tabular}{c||ccc }
 & $SU(N)$ & $U_V(1)$ & $U_A(1)$  \\
  \hline \hline \\
${\psi_{\{ij\}}}$& $\symm$ & $1$ & $1$  \\
 &&&\\
 $\widetilde{\psi}^{\{ij\}}$ &$\bsymm $& $-1$ & $1$ \\
 &&&\\
$G_{\mu}$ &{\rm Adj} & $0$ & $0$    \\
\end{tabular}
\end{minipage}
 \hskip 2cm \begin{minipage}{2.2 in}
\begin{tabular}{c||ccc}
 & $SU(N)$ & $U_V(1)$ & $U_A(1)$  \\
  \hline \hline \\
$\psi_{[ij]}$& $\asymm$ & $1$ & $1$  \\
 &&&\\
 $\widetilde{\psi}^{[ij]}$ &$\basymm $& $-1$ & $1$ \\
 &&&\\
$G_{\mu}$ &{\rm Adj} & $0$ & $0$    \\
\end{tabular}
\end{minipage}
\end{center}
\caption{The table summarizes the perturbative spectrum of the (anti) symmetric orientifold theories in the (right) left part of the table. $\psi$
and $\widetilde{\psi}$ are two Weyl fermions, while $G_{\mu}$ indicates
the gauge bosons. $U_V(1)$ is the
conserved vector global symmetry while the $U_A(1)$ symmetry is broken by the chiral quantum anomaly.} \label{table}
\end{table}

The subtle issues of the confinement
properties stemming from the centre group symmetries and (in)equivalences at finite and large $N$ were discussed first in
Ref.~\cite{Sannino:2005sk}. The large $N$ planar equivalence  means that some of the super Yang-Mills exact results, such as the ones found in  
\cite{Shifman:1999mv,Shifman:1987ia} can be exported to
the infinite $N$ limit of the orientifold theories. For example, the orientifold
theories, at large $N$, have  an exactly calculable bifermion
condensate and an infinite number of degeneracies  in the spectrum
of color-singlet hadrons. 
Some of the finite $N$ implications of the correspondence were developed in \cite{Sannino:2003xe}. The  corrections were deduced by matching the infinite $N$ effective theory describing the lowest lying composite states associated to the fluctuations of the fermion condensate with the bosonic sector of the 
Veneziano and Yankielowicz (VY) \cite{Veneziano:1982ah} effective theory for super Yang-Mills. By considering
$1/N$ supersymmetry breaking effects, as well as the explicit ones due to the introduction of a 
finite quark mass, several predictions were made in
 \cite{Sannino:2003xe} concerning the spectrum of the low-lying mesonic states, the gluon condensate, the $\theta$-angle physics as well as the vacuum energy. In particular we computed the pseudo-scalar to scalar mass ratio at finite number of colors and at finite quark mass for the orientifold field theories.

Capitalizing on the work on orientifold field theories via their description in terms of $O'$ string theory of \cite{DiVecchia:2004hvn,DiVecchia:2005vm} the same pseudo-scalar to scalar mass ratio was discussed in \cite{Armoni:2005qr} for the fermion massless limit. I will comment  later about  the different approximations that went in this analysis including the  caveat on how to interpret the computations in the closed string sector \cite{DiVecchia:2004hvn,DiVecchia:2005vm}.

Intriguingly, as  suggested in \cite{Sannino:2005sk,Feo:2004mr}, one can flip the point of view and use QCD or $SU(N)$ with higher values of $N$ for the two-index symmetric theories to acquire information about super Yang-Mills. Here I have in mind not only the (pseudo) scalar spectrum of the theory but also higher spin and excited states \cite{Ziegler:2021nbl,DellaMorte:2023ylq}. In fact, these sectors have eluded the power of supersymmetry and are relevant for model building.  Additionally,  assuming that our suggestion, i.e. the non-perturbative dynamics of chiral gauge theories reduces in the infrared to the one of a strongly coupled vector core augmented by a non-interacting chiral one, holds true one can also determine the non-perturbative spectrum of chiral gauge theories.  

Recent lattice simulations for ordinary one flavour QCD \cite{Ziegler:2021nbl,Jaeger:2022ypq,DellaMorte:2023ylq,DellaMorte:2023sdz} provide  results that can be directly compared with the original predictions \cite{Sannino:2003xe}. Of course, this is not the first time that one flavor QCD has been studied numerically. In fact, general features of the theory were discussed in \cite{Creutz:2006ts}. The quark condensate  was computed in \cite{DeGrand:2006uy} by comparing the density of low-lying eigenvalues of the
overlap Dirac operator to predictions from Random Matrix
Theory~\cite{Leutwyler:1992yt,Shuryak:1992pi}. The results are in agreement  with the
predictions for the gluino condensate obtained in
\cite{Armoni:2003yv} although sizable $1/N$ corrections are expected \cite{Sannino:2003xe}.  Using Wilson fermions the authors of reference \cite{Farchioni:2007dw}
had provided a preliminary investigation of the low-lying mesonic spectrum of one-flavor QCD. The latter was improved in 
in \cite{Ziegler:2021nbl,Jaeger:2022ypq,DellaMorte:2023ylq,DellaMorte:2023sdz}  via finer lattice spacing, larger
volumes and a tree-level improved fermionic action. Recently the glueball spectrum has been discussed in \cite{Athenodorou:2023xdi}.  Reducing the numbers of colors to two colors the one flavor case was investigated in 
 \cite{Francis:2018xjd}.  This theory is phenomenologically interesting as composite Dark Matter model. Due to the fact that the fundamental representation of $SU(2)$ is pseudo-real the global $U(1)$ vector symmetry is enhanced to $SU(2)$. Interestingly, 
 the dark matter model envisioned in   \cite{Francis:2018xjd} features a mass-gap with vector mesons being the lightest triplet of the enhanced $SU(2)$
global symmetry. The latter is a spectral feature shared with the dark matter model based also on an $SU(2)$ gauge theory but featuring scalar
quarks~\cite{Hambye:2009fg}. The $SU(2)$ theory with one Dirac adjoint fermion  was investigated, adopting the Wilson Dirac operator, in ~\cite{Athenodorou:2021wom} to elucidate the onset of the conformal window  for higher dimensional representations \cite{Sannino:2004qp,Dietrich:2006cm}. The first study of composite dark matter via lattice study appeared in \cite{Lewis:2011zb} for $SU(2)$ with two Dirac fermions. In fact, this theory is one of the most revealing and central a minimal template for composite dynamics for beyond standard model physics as summarized in \cite{Cacciapaglia:2020kgq}.  For a detailed discussion about the methodologies used on the lattice and their limits the reader will find a detailed account in \cite{Ziegler:2021nbl,Jaeger:2022ypq,DellaMorte:2023ylq,DellaMorte:2023sdz}. The issues related to the use of Wilson fermions are further discussed in ~\cite{Edwards:1997sp,Edwards:1998sh,Akemann:2010em,Mohler:2020txx,Bergner:2011zp}. I further recall that earlier  investigations of orientifold
theories~\cite{Lucini:2010kj,Armoni:2008nq} were performed via quenched approximations avoiding the sign problem but also naturally suppressing fermion loops which is more in line with the 't Hooft large $N$ limit. We complete our excursion in lattice land by recalling that numerical simulations of supersymmetric gauge theories have a long history \cite{Catterall:2009it} with a recent review  \cite{Schaich:2022xgy} discussing the status and open challenges.   
 
 The remainder  is organized as follows.  In Section~\ref{EOT}
I review the effective orientifold lagrangian of \cite{Sannino:2003xe} and briefly summarize the rationale behind its construction. I then  report the  $1/N$ predictions of the vacuum and spectral properties of the orientifold field theories.  Retracing the steps of \cite{Sannino:2003xe}, I write the physical quantities directly in terms of the trace and axial anomalies in units of the super Yang-Mills ones. The rewriting not only allows to better visualize the origin of the finite $N$ deformations from super Yang-Mills but also to arrive at an improved  version of the pseudo-scalar to scalar mass ratio which, of course, returns the original result  \cite{Sannino:2003xe} when restricting it to the $1/N$ corrections.  The leading $1/N$ and improved predictions are then  confronted with the lattice results of \cite{DellaMorte:2023ylq} showing the remarkable fact that the improved results are about one standard deviation away from  the lattice central result.  By comparing with the leading $1/N$ result I estimate the size of the $1/N^2$ corrections.   Section~\ref{EOTM} reviews the impact of  a small fermion mass operator on the spectrum. We argue that by testing the mass dependence of the pseudoscalar to scalar ratio constitutes an independent direct test of the infinite $N$  supersymmetric connection. This is so since, at leading order in the fermion mass, the corrections are protected by supersymmetry because  the mass operator is a soft susy breaking one \cite{Masiero:1984ss}. 
In section~\ref{comment} we comment on the string theory inspired computation of the pseudoscalar to scalar mass ratio at zero fermion mass. We then move up in the number of flavors and step away from super Yang Mills. Having learnt already so much about the alternative large $N$ limit we are ready to review its impact on non-perturbative scattering dynamics of two-index symmetric representations with more than one Dirac flavor \cite{Kiritsis:1989ge,Sannino:2007yp}.  Here we finally address Weinberg's point by showing that the CR extrapolation more effectively unitarize   $\pi \pi$ scattering than the 't Hooft limit. Additionally, we will see that tetraquark mesonic states appear already at  leading order in $N$ in the CR limit.  We offer our conclusions in section~\ref{summary}.

\section{Spectrum and vacuum properties via the Effective Orientifold Theory}
\label{EOT}
Effective Lagrangians describing the vacuum properties of strongly coupled gauge theories respective the underlying trace and axial variations have a long history \cite{Schechter:1980ak,Rosenzweig:1979ay,DiVecchia:1980yfw,Kawarabayashi:1980dp,Migdal:1982jp,Cornwall:1983zb}.
 Here we will consider the effective Lagrangian for the finite $N$ orientifold field theory constructed in \cite{Sannino:2003xe} which reads: 
\beq
 {\cal L}_{\rm eff}=f(N)\left\{
\frac{1}{\alpha}\left(\varphi\, \bar\varphi \right)^{-2/3}\,
\partial_\mu\bar\varphi\,\partial^\mu\varphi-\frac{4\alpha}{9}\,
\left(\varphi\, \bar\varphi \right)^{2/3}\,\left(
\ln\bar\Phi\,\ln\Phi - b \right)\right\}\, .
\label{fnocomponent}
\eeq 
The field $\varphi$ encodes information about the fermion bilinear 
\begin{eqnarray}
\varphi= -\frac{3}{32\pi^2\, N}\,
 \widetilde{\psi}^{\alpha,[i,j]} \psi_{\alpha,[i,j]} \ .
\label{phi1}
\end{eqnarray}
We have  set the infinite $N$ gluino condensate scale $\Lambda =1$ to keep the notation light but it will re-inserted later when discussing the physical spectrum. $b \sim {\cal O}(1/N)$ is a numerical (real) positive
parameter as argued in \cite{Sannino:2003xe} by comparing to the QCD gluon condensate and vacuum energy\footnote{This is possible since for $N=3$ where the two-index antisymmetric theory is QCD.},  $f(N) \rightarrow N^2$ at $N \rightarrow\infty$ and 
\beq \Phi =
\varphi^{1+\epsilon_1}\,\bar\varphi^{-\epsilon_2}\,,\qquad
\bar\Phi = \bar\varphi^{1+\epsilon_1}\,\varphi^{-\epsilon_2}\,,
\label{newfields} \eeq where $\epsilon_{1,2}$ are 
subleading in $1/N$ parameters allowing for distinct finite $N$ corrections stemming from the scale  and chiral anomalies. In fact, varying the effective oreintifold actions under scale  $\varphi \to (1+3\gamma)\varphi $
and chiral $\varphi \to (1+2i \gamma)\varphi $ transformations with respect to the real parameter $\gamma$
one obtains the following scale and chiral transformation at the effective action level: 
\begin{eqnarray}
\delta {\cal S}_{\rm eff}^{\rm scale}&=& \int d^4x\left\{-4\, \frac{\alpha\,
f}{3}\left(\varphi\, \bar\varphi \right)^{2/3}\left( 1+\epsilon_1
-\epsilon_2\right)
\left(\ln\bar\Phi +\ln \Phi\right)\right\}\,, \\[4mm]
\delta {\cal S}_{\rm eff}^{\rm chiral}&=& \int d^4 x\left\{-8i\,\, \frac{\alpha\,
f}{9}\left(\varphi\, \bar\varphi \right)^{2/3}\left( 1+\epsilon_1
+\epsilon_2\right) \left(\ln\bar\Phi -\ln \Phi\right)\right\}\,. 
\label{variat}
\end{eqnarray}
 To deduce the epsilon parameters we need to compare them to the    trace and axial anomalies at the fundamental lagrangian level:
\begin{eqnarray}
\vartheta^{\mu}_{\mu} &=& \frac{\beta_{O\pm}(a)}{a^2} \frac{1}{32\pi^2}\, G_{\mu\nu}^a {G}^{a ,\, \mu\nu}
\\[3mm]
\partial^{\mu} J_{\mu}
&=& 
\left[N \pm 2\right]\frac{1}{16\pi^2}\, G_{\mu\nu}^a {\tilde{G}}^{a
,\, \mu\nu} \ . \label{axial}
\end{eqnarray}
 Here $\beta_{O\pm}$ are the beta functions for the orientifold field theories with the generic beta function defined, in perturbation theory, as: 
\begin{equation}
\frac{d\, a}{d \ln \mu^2}\equiv\beta_{O\pm}(a) = - \beta_0 a^2 - \beta_1 a^3 - \beta_2 a^4 \cdots \ ,  \quad {\rm with} \quad a=g^2/(4\pi)^2 \ .
\end{equation}
Additionally by comparison between the underlying and effective anomaly variations we have \cite{Sannino:2003xe}:
\begin{eqnarray}
G_{\mu\nu}^a {G}^{a ,\, \mu\nu} &\propto& -N\left(\varphi\, \bar\varphi
\right)^{2/3}\left(\ln\bar\Phi +\ln \Phi\right)\,,\nonumber\\[3mm]
G_{\mu\nu}^a \tilde{G}^{a ,\, \mu\nu} &\propto& -N\,i\, \left(\varphi\, \bar\varphi
\right)^{2/3}\left(\ln\bar\Phi -\ln \Phi\right)\,.
\label{indenti}
\end{eqnarray}
 Once properly normalized to the super Yang-Mills limit we have: 
\begin{equation}
1+\epsilon_1+\epsilon_2  \equiv {\chi} =   \frac{N \pm 2}{N} \ ,\qquad 1+\epsilon_1-\epsilon_2  \equiv {\cal T}  =\frac{\beta_{O\pm}(a)}{\beta_{SYM}(a)}= -\frac{1}{3N}\frac{\beta_{O\pm}(a)}{ a^2} \ ,
\label{fixingepsilon}
\end{equation}
 yielding: 
\begin{equation}
2\epsilon_2 = \chi - {\cal T} \ , \qquad \epsilon_1 = 1 - \frac{\chi + {\cal T}}{2} \ .
\end{equation}
In the above equations and the following ones $\pm$ refers respectively to the two-index symmetric and antisymmetric representation. 
By construction both epsilons start at the order $1/N$ when moving away from super Yang-Mills.  
Once minimized the effective orientifold action returns the following vacuum energy density: 
\beq V_{\rm min}={\cal E}_{\rm vac} = -\frac{4\alpha\,
f}{9}\, b + O(N^0)\,.
\label{vmin}
 \eeq
 As we mentioned earlier the sign of $ b$ is consistent, for the two-index antisymmetric orientifold theory, with the fact that for QCD with three light flavors the gluon condensate is  positive and the
vacuum energy density is negative \cite{Shifman:1978bx}. 
One flavor QCD  is obtained by 
decoupling two light quarks affecting the absolute value
of the gluon condensate but  not its sign
\cite{Novikov:1981xi}.  For further  arguments related to the sign of $b$ associated, for example, to pure gluon-dynamics we refer to \cite{Sannino:2003xe}. 


Approaching super Yang-Mills when increasing the number of colors leads to spectral degenaracies for the orientifold theories dictated by super symmetry and  reflected in the effective orientifold Lagrangian  by the  degeneracy between the  scalar $\sigma$ and the pseudoscalar $\eta^{\prime}$
mesons. Studying the spectrum of fluctuations related to $\varphi$  as $\displaystyle{\varphi = \langle \varphi \rangle \left(1
 + c\,h\right)}$ with $\displaystyle{
 h=\frac{1}{\sqrt{2}} \left(\sigma + i\,\eta^{\prime}
\right)}$ and $\displaystyle{c^2= \frac{\alpha}{f} \, |\langle  \varphi \rangle|^{-\frac{2}{3}}}$ we obtained in \cite{Sannino:2003xe}: 
\begin{eqnarray}
 M_{\sigma} &=&\frac{2\alpha}{3}\, {\Lambda} \, (1+\frac{2}{9}b) \,
\left[1 + \epsilon_1 - \epsilon_2+ \frac{4}{9}b \right]
=\frac{2\alpha}{3}\, {\Lambda}\, (1+\frac{2}{9}b) \,\left[{\cal T}+
\frac{4}{9} b\right], 
\nonumber \\[3mm]
M_{\eta^{\prime}} &=&\frac{2\alpha}{3}\, {\Lambda} \, (1+\frac{2}{9}b) \,\left[1 +
\epsilon_1  + \epsilon_2 \right] 
=\frac{2\alpha}{3}\, {\Lambda}\,\, (1+\frac{2}{9}b) \,\chi \ ,  
\label{spectrum}
 \end{eqnarray}
where in the last equality we expanded at leading order in $1/N$. In fact it is not necessary to make this expansion when considering the ratio of the masses. At infinite number of colors the $b$ coefficient vanishes and ${\cal T}= \chi = 1$  and we recover   the supersymmetric limit $\displaystyle{M_\sigma = M_\eta^{\prime} = \frac{2\alpha}{3} \Lambda}$. The vacuum parameter $b$ beyond the overall redefinition of $\Lambda$ for both physical states it further affects directly the mass of the scalar state since the latter has the same quantum numbers of the vacuum, however it does not contribute to the pseudo-scalar state because the orientifold theory is parity invariant.

The finite $N$ ratio of the pseudoscalar to scalar
mass is
 \begin{eqnarray}
\frac{M_{\eta^{\prime}}}{M_{\sigma}} =  \frac{\chi} {{\cal T} + \frac{4}{9}b}  =  \frac{1 \pm {2}/{N}}{ 
{\beta_{O\pm}}/{\beta_{SYM}} + \frac{4}{9} b}  \leq   \frac{1 \pm {2}/{N}}{ {\beta_{O\pm}}/{\beta_{SYM}}  }\ .\label{spectrum-ration}
\end{eqnarray} 
Considering only the leading $1/N$ corrections we recover the result of  \cite{Sannino:2003xe} 
 \begin{eqnarray}
\frac{M_{\eta^{\prime}}}{M_{\sigma}} =  {1 + 2\epsilon_2 - \frac{4}{9} b} + {\cal O}(N^{-2}) = 1 + \chi - {\cal T} - \frac{4}{9}b  + {\cal O}(N^{-2}) \ ,
  \label{spectrum-perturbative-ratio}
\end{eqnarray} 
where in the last equality we keep only the leading $1/N$ corrections in the  $\chi - {\cal T}$ term. Clearly equation (\ref{spectrum-perturbative-ratio}) holds in the vicinity of $N\rightarrow \infty$ while we expect equation (\ref{spectrum-ration}) to better capture the general features of this ratio for smaller values of $N$.

To be predictive one can now estimate the ratio $\beta_{O\pm}/\beta_{SYM}$. In \cite{Sannino:2003xe}  we employed the one loop  beta function coefficients. The rationale for this choice is that, in the Wilsonian regularization scheme used to  write the VY effective action for  super Yang-Mills the beta function is one loop exact. In turns, this ensures that the effective potential of the effective supersymmetric action is holomorhpic.  Therefore, when going away from the supersymmetric limit it is natural to start with the one loop coefficient of the orientifold beta functions that captures in a simple manner deviations from holomorphicity yielding
\begin{equation}
{\cal T} = \frac{\beta_{O\pm}} {\beta_{SYM} } = 1  \mp \frac{4}{9N}   \qquad {\rm one~loop} \ . 
\label{one-loop-test}
\end{equation}
Expanding around the infinite $N$ result and retaining only the $1/N$ corrections one recovers the result of \cite{Sannino:2003xe} which is: 
 \begin{eqnarray}
 \frac{M_{\eta^{\prime}}}{M_{\sigma}}\Big|_{O\pm} =   1 + \chi - {\cal T} - \frac{4}{9}b  + {\cal O}(N^{-2})  = 1 \pm \frac{22}{9N} - \frac{4}{9}b  + {\cal O}(N^{-2})\ .
  \label{spectrum-explicit-perturbative}
\end{eqnarray} 
Needless to say, this result is valid only in the asymptotically large number of colors and direct comparison for small $N$ are subject to large corrections. If one insists in trying to access smaller values of $N$ it should be better to compare  to   (\ref{spectrum-ration}): 
 \begin{eqnarray}
\frac{M_{\eta^{\prime}}}{M_{\sigma}}\Big|_{O\pm} =  \frac{\chi} {{\cal T} + \frac{4}{9}b}  \,  \leq   \frac{1 \pm {2}/{N}}{ 1 \mp \frac{4}{9N}  }\ , \label{spectrum-ration-explicit} 
\end{eqnarray}
which, of course, recovers the expected large $N$ subleading terms of \cite{Sannino:2003xe} and  better captures finite $N$ corrections, as we shall momentarily see, when comparing the prediction to lattice data \cite{Ziegler:2021nbl,Jaeger:2022ypq,DellaMorte:2023ylq,DellaMorte:2023sdz}. I  term the  ratio in  \eqref{spectrum-ration-explicit} the  finite $N$ improved result.    

\section{Comparing to lattice results and the size of the $1/N^2$ corrections}

\begin{figure}[h!]
    \centering
    \includegraphics[width=.75\linewidth]{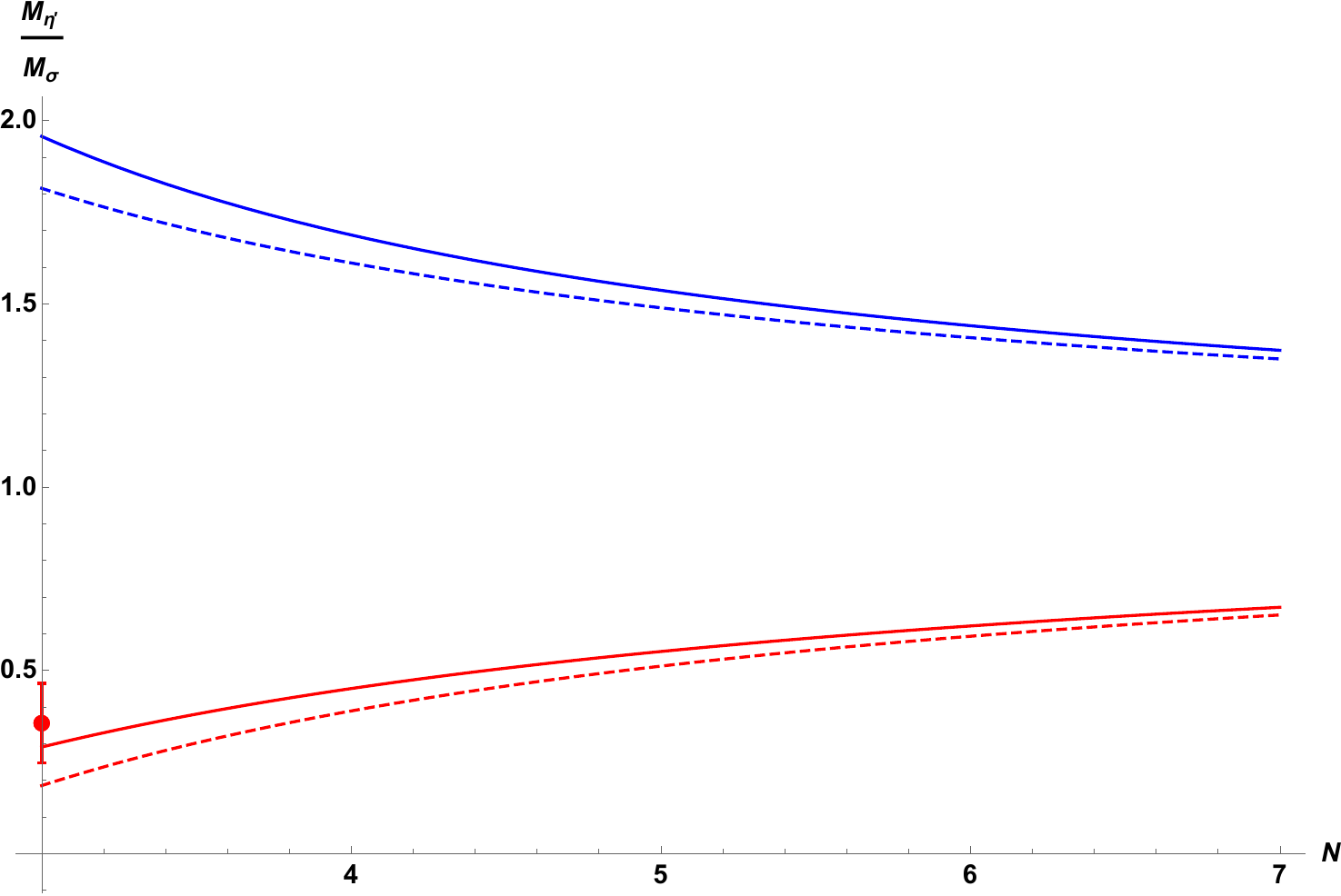}
    \caption{Predictions for the pseudoscalar to scalar mass ratio as function of the number of colors $N$ for the two-index symmetric, upper blue solid and dashed curves, and two-index antisymmetric as lower red solid and dashed curves.  The solid lines are due to \eqref{spectrum-ration-explicit} while the dashed lines come from \eqref{spectrum-ration}. The lattice result, for one-flavor QCD, is shown at two sigma confidence level. 
}
    \label{fig:enter-label}  
\end{figure}

For three colors the two-index antisymmetric theory is one flavor QCD for which the  $1/N$ and the improved results give respectively 
 \begin{equation}
\frac{M_{\eta^{\prime}}}{M_{\sigma}}\Big|_{\rm QCD-leading} 
\leq 0.185 ~{\rm up~to~}{\cal O}(N^{-2})   \quad  {\rm and} \quad \frac{M_{\eta^{\prime}}}{M_{\sigma}}\Big|_{\rm QCD-improved} 
\leq 0.290 \ . 
\label{QCD} 
\end{equation}
The predictions should be compared to the lattice value $0.356(54)$ computed in \cite{DellaMorte:2023ylq,DellaMorte:2023sdz}. The latter is compatible with the improved result within two sigma confidence level. In figure~\ref{fig:enter-label} I plot our predictions for the pseudoscalar to scalar mass ratio as function of the number of colors for both, the two-index symmetric (blue curves) and antisymmetric case (red curves). The dashed curves correspond to the leading $1/N$  corrections while the solid ones are for the improved version. The lattice result for one-flavor QCD is shown in the plot for $N=3$ at the two sigma confidence level.

Given the remarkable success it is tempting to assume  the  improved expression 
 \begin{eqnarray}
\frac{M_{\eta^{\prime}}}{M_{\sigma}} =  \frac{\chi} {{\cal T} + \frac{4}{9}b}  =  \frac{1 \pm {2}/{N}}{ 
{\beta_{O\pm}}/{\beta_{SYM}} + \frac{4}{9} b}  \ ,\label{spectrum-general}
\end{eqnarray} 
to be valid for any $N$. After all, the numerator coefficient, that stems from the axial anomaly, is all-order exact and we have seen that  mild assumptions on the ratio of beta functions  leads to surprisingly consistent results with first principle lattice simulations even for small number of colors while naturally interpolating with the leading $1/N$ corrections. 

 It is natural to ask what happens if one were to depart from the Wilsonian scheme for the beta functions. One way to illustrate this is to consider the universal two-loop beta ratio expanded in $1/N$ which yields: 
 \begin{equation}
{\cal T} = \frac{\beta_{O\pm}}{\beta_{SYM}} =  1 \mp \frac{4}{9 N} \frac{ 1+\frac{19}{4} \lambda }  {1+ \lambda}  \quad {\rm with} \quad  \lambda = \frac{N g^2}{8\pi^2} \ .
\end{equation}
Here $\lambda$ is the 't Hooft coupling and one can easily check that the result is consistent with the all-order NVSZ super Yang-Mills  beta function \cite{Novikov:1983uc,Shifman:1986zi}. One can always find a scheme were the two-loop is all there is. In this case the largest modification occurs for the largest value of $\lambda$  giving ${\cal T} = 1\mp 19/(9N)$ and an upper limit for the QCD ration of 0.196, stemming from \eqref{spectrum-ration-explicit}, which is slightly higher than the leading $N$ one. This speaks again  in favour of the robustness of the results.

We now use the lattice results to estimate the size of the $1/N^2$ by extending  \eqref{spectrum-explicit-perturbative} to   
 \begin{eqnarray}
 \frac{M_{\eta^{\prime}}}{M_{\sigma}}\Big|_{O\pm} =   1 + \chi - {\cal T} - \frac{4}{9}b  + {\cal O}(N^{-2})  = 1 \pm \frac{22}{9N} - \frac{4}{9}b  + 
 \frac{\kappa}{N^2}+{\cal O}(N^{-3})\ , 
  \label{spectrum-explicit-perturbative}
\end{eqnarray} 
and obtain $\kappa \geq 1.54$ with the equality holding for $b=0$. For a reasonable positive non vanishing value of $b \approx 1/N$ we find $\kappa \approx 2.87$.

\section{The quark mass weighs in}
\label{EOTM}
In \cite{Sannino:2003xe} we followed    Masiero and Veneziano \cite{Masiero:1984ss} in adding a 
supersymmetry breaking term  stemming from the gluino mass $m$ translating into adding in the initial  Lagrangian the last operator below:
\beq f(N)\left\{ \frac{1}{\alpha}\left(\varphi\, \bar\varphi
\right)^{-2/3}\,
\partial_\mu\bar\varphi\,\partial^\mu\varphi-\frac{4\alpha}{9}\,
\left(\varphi\, \bar\varphi \right)^{2/3}\,\left(
\ln\bar\Phi\,\ln\Phi - b \right)\right\} + \frac{4\,
m}{3\lambda}\, N^2 \left(\varphi + \bar{\varphi}\right)\ .
\label{fnocomponent+mass} \eeq
 This yields the following vacuum energy: 
\begin{eqnarray}
{\cal E}_{\rm vac}=V_{\rm min} = -\frac{4\alpha f}{9} b
\Lambda^4 - \frac{8N^2}{3\lambda} m\Lambda^3 +{
O}\left(m^2,N^{0},mN\right) \ .
\end{eqnarray}
and the leading $1/N$ and $m$ spectrum for the ratio of the pseudoscalar to scalar mass one has: 

 \begin{eqnarray}
\frac{M_{\eta^{\prime}}}{M_{\sigma}} = 
 1 -\frac{22}{9N} -\frac{4}{9}b -  \frac{m}{\alpha \lambda \Lambda}+
{O}(m^2,N^{-2},mN^{-1}) \ .\label{spectrum-ration-Nm}
\end{eqnarray}
The improved finite $N$ but small fermion mass version is
\begin{eqnarray}
\frac{M_{\eta^{\prime}}}{M_{\sigma}}\Big|_{\rm improved} = 
 \frac{1 \pm {2}/{N}}{ 
{\beta_{O\pm}}/{\beta_{SYM}} + \frac{4}{9} b} -  \frac{m}{\alpha \lambda \Lambda}+
{O}(m^2) \ .\label{spectrum-ration-Improved}
\end{eqnarray}

The gluon condensate becomes:
\begin{eqnarray}
 \frac{\langle G^a_{\mu\nu}G^{a,\mu\nu} \rangle}{64\pi^2} =
\frac{4\, N \, m}{3\lambda}\, \Lambda^3 +
\frac{8}{27}\alpha\,N\,b\Lambda^4+ { O}\left(m^2,N^{-1},mN^0
\right)\ .
\end{eqnarray} 

These results show that the contribution of the fermion mass
reinforces the effect of the finite $N$ contribution in decreasing the ratio of the pseudo-scalar to scalar masses.

The $\theta$-angle dependence of the vacuum energy for the
fermions in the two-index antisymmetric representation of the
gauge group is
\begin{eqnarray} {\cal E}_{\rm vac} =
\frac{8N^2}{3\,\lambda}\, m \Lambda^3\, {\rm
min}_{k}\left\{-\cos\left[\frac{\theta + 2\pi\,k
}{N-2}\right]\right\}  -\frac{4\alpha f}{9} \, b\Lambda^4 \
.\end{eqnarray}
The $N-2$-fold vacuum degeneracy is lifted due to the presence of
a mass term in the theory, yielding a unique vacuum.

\section{Comments on a string theory inspired perspective}
\label{comment}

Using a string inspired approach the authors of reference \cite{Armoni:2005qr} provided an alternative estimate for the pseudo-scalar to scalar mass ratio above by arguing in favour of setting ${\cal T} + \frac{4}{9}b \approx 1$ while keeping $\chi = 1\pm 2/N$. This choice   corresponds, in the effective Lagrangian approach, to have the scalar mass fixed at the SUSY value and therefore only the axial anomaly is responsible for the  finite $N$ corrections, arriving at
\begin{equation}\frac{M_{\eta^{\prime}}}{M_{\sigma}}\Big|_{\rm String} \approx \chi = 1\pm \frac{2}{N} \ . \end{equation} 
I will now briefly review the assumptions made in \cite{Armoni:2005qr} and highlight the various caveats. The first intuition is that one can extract information about orientifold field theories via their description in terms of $O'$ string theory of \cite{DiVecchia:2004hvn,DiVecchia:2005vm}. By determining the string annulus partition function in presence of external gauge fields for the open and close string sectors in \cite{DiVecchia:2004hvn,DiVecchia:2005vm} the coefficients for the axial and trace anomalies were determined. The axial coefficients were computed to be identical for both the closed and opened string sectors naturally matching the exact field theoretical result while the trace anomaly coefficients differ, with the open strings yielding the one-loop result for orientifold field theories and the closed sector giving the supersymmetric result. Naturally, the authors of \cite{DiVecchia:2004hvn,DiVecchia:2005vm}  attributed the discrepancy, in computing the coefficient of the trace anomaly, in the open and closed string sectors to the missing of threshold corrections in the closed sector. In fact, both sectors should return the perturbative field theoretical result. Despite this warning, the authors  in \cite{Armoni:2005qr}   elevated the incomplete closed string result to a genuine non-perturbative prediction of trace anomaly for the orientifold field theory. A second assumption is that the pure (pseudo) glueball  point function is saturated by the (pseudo) scalar state assuming that these are the lightest states. This assumption is partially justified by the studies in \cite{Merlatti:2004df,Feo:2004mr}. In these papers the effective Lagrangians included  besides the (pseudo) scalar states made by fermions  also the  (pseudo) glueball states and computed their mixing.   The last assumption \cite{Armoni:2005qr} is to neglect extra ${\cal O}(1/N)$ corrections in any other existing parameter entering the ratio of the two-point functions except for the   coefficient of the axial anomaly. Overall, seen from the effective orientifold theory computation, the approximations above amount at considering only the corrections for the pseudoscalar state (stemming from axial anomaly) while the scalar state remains unperturbed from the super Yang-Mills limit which would be surprising. Nevertheless, given that one of the strongest assumptions is associated to the string theory description, it would be interesting to have a complete computation for the closed sector of $O'$ string theory. Of course, a direct comparison with the lattice data, even if they superficially seem to be a better fit  still begs the question of why the $1/N^2$ corrections should be so much suppressed.

\section{Tetraquark mesons and $\pi \pi$ scattering at large $N$, explained}
\label{unity}

I can now comment on Weinberg's work \cite{Weinberg:2013cfa} about large $N$ meson-meson scattering and tetraquarks. These two quantities are naturally related since tetraquarks can appear as intermediate states in the $\pi \pi$ scattering amplitude. Upon analyzing the tetraquark contribution to meson-meson scattering and their propagators, Coleman \cite{Coleman_1985} correctly argues that, in the leading $N$ 't Hooft limit, these states do not appear in meson-meson scattering and furthermore the connected component of their propagator is also suppressed. In Coleman's own words: {\it In the large $N$ limit, quadrilinear make meson pairs and nothing else}.  Weinberg challenges this statement and writes: {\it But is this justified?}. Rather than answering the question at the scattering amplitude level he considers the decay width instead.  Weinberg furthermore argues that Coleman's conclusion could be reasonable if the decay width into ordinary mesons grows with $N$, justifying why a tetraquark may not be observable  as a distinct particle. He then argues that the width decreases with $N$ instead\footnote{As a sign of phenomenological success Weinberg mentions the small with of the $f_0(980)$ tetraquark state. However, for this state, the width is small because of phase space since it would rather decay into $K{\bar{K}}$ as shown some time ago in \cite{Harada:1995dc}.}.  Weinberg interprets his result as a way to justify the presence of tetraquarks even in the 't Hooft limit. However, this is not entirely correct, since the mass of the tetrquarks could be not leading in $N$ and some tetraquarks can still remain broad, like the $f_0(500)$. 
 
\begin{figure}[htbp]
\centering

\tikzset{every picture/.style={line width=0.7pt}} 

\scalebox{0.95}{\begin{tikzpicture}[x=0.75pt,y=0.75pt,yscale=-1,xscale=1]

\draw  [fill={rgb, 255:red, 0; green, 0; blue, 0 }  ,fill opacity=1 ] (186.63,194.32) -- (170.19,200.17) -- (169.51,190.94) -- cycle ;
\draw  [fill={rgb, 255:red, 0; green, 0; blue, 0 }  ,fill opacity=1 ] (265.61,193.38) -- (249.12,199.12) -- (248.5,189.88) -- cycle ;
\draw    (145,195.35) -- (288,193.65) ;
\draw   (216.01,194.12) .. controls (214.04,195.15) and (212.17,196.14) .. (212.17,197.27) .. controls (212.18,198.41) and (214.06,199.38) .. (216.04,200.4) .. controls (218.01,201.41) and (219.89,202.39) .. (219.9,203.52) .. controls (219.9,204.66) and (218.03,205.65) .. (216.06,206.68) .. controls (214.1,207.71) and (212.22,208.7) .. (212.23,209.83) .. controls (212.23,210.97) and (214.11,211.94) .. (216.09,212.96) .. controls (218.07,213.97) and (219.95,214.95) .. (219.95,216.08) .. controls (219.96,217.22) and (218.09,218.21) .. (216.12,219.24) .. controls (214.15,220.27) and (212.28,221.26) .. (212.28,222.39) .. controls (212.29,223.53) and (214.17,224.5) .. (216.15,225.52) .. controls (218.12,226.53) and (220,227.51) .. (220.01,228.64) .. controls (220.01,229.78) and (218.14,230.77) .. (216.17,231.8) .. controls (214.21,232.83) and (212.33,233.82) .. (212.34,234.95) .. controls (212.34,236.09) and (214.22,237.06) .. (216.2,238.08) .. controls (218.18,239.09) and (220.06,240.07) .. (220.06,241.2) .. controls (220.07,242.34) and (218.2,243.33) .. (216.23,244.36) .. controls (214.26,245.39) and (212.39,246.38) .. (212.39,247.51) .. controls (212.4,248.65) and (214.28,249.62) .. (216.26,250.64) .. controls (218.23,251.66) and (220.11,252.63) .. (220.12,253.76) .. controls (220.12,254.9) and (218.25,255.89) .. (216.28,256.92) .. controls (214.32,257.95) and (212.44,258.94) .. (212.45,260.07) .. controls (212.45,261.21) and (214.33,262.18) .. (216.31,263.2) .. controls (217.14,263.63) and (217.96,264.05) .. (218.62,264.48) ;

\draw   (300,223) -- (312.5,215) -- (312.5,219) -- (337.5,219) -- (337.5,215) -- (350,223) -- (337.5,231) -- (337.5,227) -- (312.5,227) -- (312.5,231) -- cycle ;
\draw  [fill={rgb, 255:red, 0; green, 0; blue, 0 }  ,fill opacity=1 ] (410.22,190.47) -- (395.29,194.57) -- (395.14,186.87) -- cycle ;
\draw  [fill={rgb, 255:red, 0; green, 0; blue, 0 }  ,fill opacity=1 ] (470.37,189.67) -- (455.42,193.71) -- (455.31,186) -- cycle ;
\draw    (365.5,191.1) -- (493,189.53) ;
\draw    (368.17,203.85) -- (426.13,204.59) ;
\draw    (426.79,256.17) -- (425.82,204.12) ;
\draw    (440.08,255.42) -- (439.19,205.37) ;
\draw    (439.02,204.35) -- (492.53,203.09) ;
\draw  [fill={rgb, 255:red, 0; green, 0; blue, 0 }  ,fill opacity=1 ] (410.22,204.21) -- (395.29,208.32) -- (395.14,200.61) -- cycle ;
\draw  [fill={rgb, 255:red, 0; green, 0; blue, 0 }  ,fill opacity=1 ] (471.26,203.83) -- (456.32,207.87) -- (456.2,200.16) -- cycle ;
\draw  [fill={rgb, 255:red, 0; green, 0; blue, 0 }  ,fill opacity=1 ] (426.28,243.81) -- (421.39,230.02) -- (429.63,229.62) -- cycle ;
\draw  [fill={rgb, 255:red, 0; green, 0; blue, 0 }  ,fill opacity=1 ] (440.06,229.81) -- (444.39,243.77) -- (436.13,243.87) -- cycle ;
\end{tikzpicture}
}
\caption[]
{two-index fermion - gluon vertex.} \label{FigA}
\end{figure}
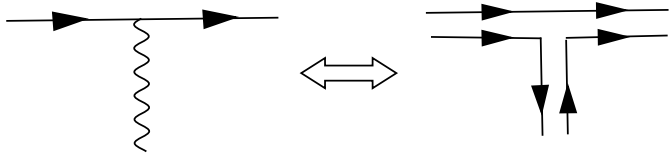

In the following I will show, at the  meson-meson scattering amplitude level, that the alternative large $N$ limit is better suited than the 't Hooft extrapolation to address the issues above. 
Because I wish to discuss phenomenologically relevant scattering amplitudes I add light flavors. Although, in this case,  the large $N$ theories depart from the supersymmetric Yang-Mills limit, one can still employ  the CR large $N$ to access information complementary to the 't Hooft limit  \cite{Kiritsis:1989ge,Sannino:2007yp}.  A noteworthy example is the celebrated $\pi\pi$ scattering amplitude \cite{Sannino:2007yp}.  To this end I recall that for quarks in the two-index representation the large $N$ diagram for the quark-quark gluon interaction is  the one in Fig.~\ref{FigA}. 

\begin{figure}[htbp]
\centering 

\tikzset{every picture/.style={line width=0.75pt}} 

\scalebox{0.50}{\begin{tikzpicture}[x=0.75pt,y=0.75pt,yscale=-1,xscale=1]

\draw   (185.5,205.5) .. controls (185.5,150.82) and (229.82,106.5) .. (284.5,106.5) .. controls (339.18,106.5) and (383.5,150.82) .. (383.5,205.5) .. controls (383.5,260.18) and (339.18,304.5) .. (284.5,304.5) .. controls (229.82,304.5) and (185.5,260.18) .. (185.5,205.5) -- cycle ;
\draw   (198.5,205.47) .. controls (198.52,157.97) and (237.03,119.48) .. (284.53,119.5) .. controls (332.03,119.52) and (370.52,158.03) .. (370.5,205.53) .. controls (370.48,253.03) and (331.97,291.52) .. (284.47,291.5) .. controls (236.97,291.48) and (198.48,252.97) .. (198.5,205.47) -- cycle ;
\draw  [fill={rgb, 255:red, 218; green, 24; blue, 24 }  ,fill opacity=1 ] (175,205.5) .. controls (175,199.7) and (179.7,195) .. (185.5,195) .. controls (191.3,195) and (196,199.7) .. (196,205.5) .. controls (196,211.3) and (191.3,216) .. (185.5,216) .. controls (179.7,216) and (175,211.3) .. (175,205.5) -- cycle ;
\draw  [fill={rgb, 255:red, 0; green, 0; blue, 0 }  ,fill opacity=1 ] (284.61,119.04) -- (268.37,125.44) -- (267.38,116.23) -- cycle ;
\draw  [fill={rgb, 255:red, 0; green, 0; blue, 0 }  ,fill opacity=1 ] (284.61,106.38) -- (268.12,112.12) -- (267.5,102.88) -- cycle ;

\end{tikzpicture}
}
\caption[]
{Diagram for $F_\pi$ for the two-index quark.} \label{FigB}
\end{figure}
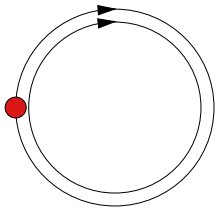

The two-index quark is represented by two lines (rather than one) with arrows pointing in the 
same direction (from which the orientifold nature is evident). The gluon, transforming according to the adjoint representation, is represented by two lines but with arrows pointing in opposite directions. The coupling constant present in the  
vertex is given by $\lambda /\sqrt{N}$, where $\lambda$ is taken to be constant and it is the 't Hooft  coupling.  In this limit the large $N$ counting for the pion decay constant $F_\pi$  can be read off from Fig.~\ref{FigB}.
          
Here the pion insertion (red dot) provides a $1/N$ normalization factor at large $N$ stemming from the finite $N$ normalization factor for the pion's wavefunction,
\begin{eqnarray}
            \frac{\sqrt{2}}{\sqrt{N(N-1)}} \ .
\label{wf}
\end{eqnarray}        
 One can now compute the leading $N$ depedence of the pion decay constant by recalling that in the loop the two oriented lines 
carry each a color index so the loop scales as $N^2$ at large $N$ and more
precisely for the antisymmetric theory we have
\begin{equation}          
 \frac{N(N-1)}{2} \ .  
\label{loop}
\end{equation}
Combining Eqs. (\ref{wf}) and (\ref{loop})
 yields the  
$F_{\pi}$ scaling in units of the three color value:
\begin{equation}
F_{\pi}^2(N)= \frac{N(N-1)}{6}F_{\pi}^2(3) \ .
\label{fpiscaling}
\end{equation}

Therefore the large $N$ scaling of $F_\pi$ is proportional to $N$ which is $\sqrt{N}$ larger than in the 't Hooft case. 

To compute the large $N$ counting for the $\pi \pi$ scattering amplitude one can employ the simplified diagram  of Fig.~\ref{FigC} with the four pion insertions.

\tikzset{every picture/.style={line width=0.75pt}} 
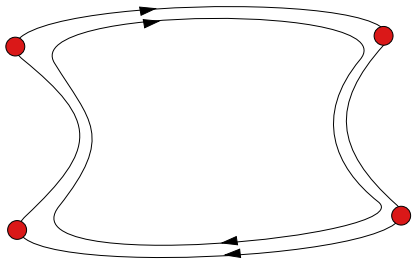
\begin{figure}[htbp]
\centering 

\scalebox{0.45}{\begin{tikzpicture}[x=0.75pt,y=0.75pt,yscale=-1,xscale=1]
\draw   (132,90) .. controls (48,21) and (591,8) .. (532,74) .. controls (473,140) and (477,190) .. (546,252) .. controls (615,314) and (55,341) .. (135,265) .. controls (215,189) and (216,159) .. (132,90) -- cycle ;
\draw  [fill={rgb, 255:red, 218; green, 24; blue, 24 }  ,fill opacity=1 ] (113,75.5) .. controls (113,69.7) and (117.7,65) .. (123.5,65) .. controls (129.3,65) and (134,69.7) .. (134,75.5) .. controls (134,81.3) and (129.3,86) .. (123.5,86) .. controls (117.7,86) and (113,81.3) .. (113,75.5) -- cycle ;
\draw  [fill={rgb, 255:red, 218; green, 24; blue, 24 }  ,fill opacity=1 ] (115,280.5) .. controls (115,274.7) and (119.7,270) .. (125.5,270) .. controls (131.3,270) and (136,274.7) .. (136,280.5) .. controls (136,286.3) and (131.3,291) .. (125.5,291) .. controls (119.7,291) and (115,286.3) .. (115,280.5) -- cycle ;
\draw  [fill={rgb, 255:red, 218; green, 24; blue, 24 }  ,fill opacity=1 ] (541,264.5) .. controls (541,258.7) and (545.7,254) .. (551.5,254) .. controls (557.3,254) and (562,258.7) .. (562,264.5) .. controls (562,270.3) and (557.3,275) .. (551.5,275) .. controls (545.7,275) and (541,270.3) .. (541,264.5) -- cycle ;
\draw  [fill={rgb, 255:red, 218; green, 24; blue, 24 }  ,fill opacity=1 ] (521.5,63.5) .. controls (521.5,57.7) and (526.2,53) .. (532,53) .. controls (537.8,53) and (542.5,57.7) .. (542.5,63.5) .. controls (542.5,69.3) and (537.8,74) .. (532,74) .. controls (526.2,74) and (521.5,69.3) .. (521.5,63.5) -- cycle ;
\draw   (167,94) .. controls (127,29) and (558,27) .. (506,90) .. controls (454,153) and (477,210) .. (526,249) .. controls (575,288) and (110,333) .. (172,254) .. controls (234,175) and (207,159) .. (167,94) -- cycle ;
\draw  [fill={rgb, 255:red, 0; green, 0; blue, 0 }  ,fill opacity=1 ] (282.79,47.06) -- (267,54.5) -- (265.42,45.38) -- cycle ;
\draw  [fill={rgb, 255:red, 0; green, 0; blue, 0 }  ,fill opacity=1 ] (352.68,295.03) -- (368.61,287.89) -- (370.02,297.04) -- cycle ;
\draw  [fill={rgb, 255:red, 0; green, 0; blue, 0 }  ,fill opacity=1 ] (278.79,33.06) -- (263,40.5) -- (261.42,31.38) -- cycle ;
\draw  [fill={rgb, 255:red, 0; green, 0; blue, 0 }  ,fill opacity=1 ] (356.14,308.43) -- (372.35,301.97) -- (373.37,311.18) -- cycle ;

\end{tikzpicture}

}

\caption[]
{Diagram for the scattering amplitude for 2 index quarks. } \label{FigC}
\end{figure}

Combining the large $N$ rules given above one arrives at \cite{Kiritsis:1989ge,Sannino:2007yp}   
\begin{equation}
A(\pi\pi \rightarrow \pi \pi)_N=\frac{6}{N(N-1)}A(\pi\pi \rightarrow \pi \pi)_{N=3},
\label{Ascaling}
\end{equation}
vanishing faster at large $N$ when compared to the 't Hooft extrapolation by an extra $1/N$ suppression factor.

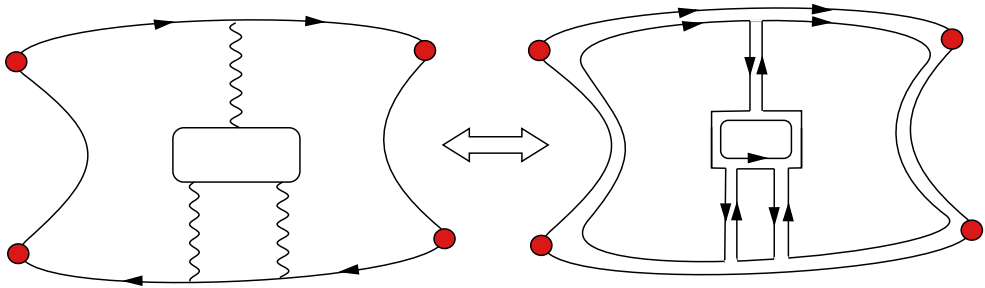
\begin{figure}[htbp]
\centering 
 
\tikzset{every picture/.style={line width=0.75pt}} 

\scalebox{0.75}{\begin{tikzpicture}[x=0.75pt,y=0.75pt,yscale=-1,xscale=1]

\draw   (18.65,134.3) .. controls (-37.29,90.97) and (324.31,82.8) .. (285.02,124.25) .. controls (245.73,165.7) and (248.4,197.1) .. (294.35,236.04) .. controls (340.29,274.98) and (-32.62,291.94) .. (20.65,244.21) .. controls (73.92,196.48) and (74.59,177.63) .. (18.65,134.3) -- cycle ;
\draw  [fill={rgb, 255:red, 218; green, 24; blue, 24 }  ,fill opacity=1 ] (6,125.19) .. controls (6,121.55) and (9.13,118.6) .. (12.99,118.6) .. controls (16.85,118.6) and (19.98,121.55) .. (19.98,125.19) .. controls (19.98,128.84) and (16.85,131.79) .. (12.99,131.79) .. controls (9.13,131.79) and (6,128.84) .. (6,125.19) -- cycle ;
\draw  [fill={rgb, 255:red, 218; green, 24; blue, 24 }  ,fill opacity=1 ] (7.33,253.94) .. controls (7.33,250.3) and (10.46,247.35) .. (14.32,247.35) .. controls (18.19,247.35) and (21.32,250.3) .. (21.32,253.94) .. controls (21.32,257.58) and (18.19,260.53) .. (14.32,260.53) .. controls (10.46,260.53) and (7.33,257.58) .. (7.33,253.94) -- cycle ;
\draw  [fill={rgb, 255:red, 218; green, 24; blue, 24 }  ,fill opacity=1 ] (291.02,243.89) .. controls (291.02,240.25) and (294.15,237.3) .. (298.01,237.3) .. controls (301.87,237.3) and (305,240.25) .. (305,243.89) .. controls (305,247.53) and (301.87,250.49) .. (298.01,250.49) .. controls (294.15,250.49) and (291.02,247.53) .. (291.02,243.89) -- cycle ;
\draw  [fill={rgb, 255:red, 218; green, 24; blue, 24 }  ,fill opacity=1 ] (278.03,117.66) .. controls (278.03,114.02) and (281.16,111.06) .. (285.02,111.06) .. controls (288.88,111.06) and (292.01,114.02) .. (292.01,117.66) .. controls (292.01,121.3) and (288.88,124.25) .. (285.02,124.25) .. controls (281.16,124.25) and (278.03,121.3) .. (278.03,117.66) -- cycle ;
\draw [color={rgb, 255:red, 255; green, 255; blue, 255 }  ,draw opacity=1 ]   (167.48,104.66) -- (181.75,105.44) ;
\draw [color={rgb, 255:red, 255; green, 255; blue, 255 }  ,draw opacity=1 ][line width=2.25]    (167.82,263.05) -- (177.14,262.42) ;
\draw [color={rgb, 255:red, 255; green, 255; blue, 255 }  ,draw opacity=1 ][line width=2.25]    (135.19,265.24) -- (143.18,263.67) ;
\draw  [fill={rgb, 255:red, 0; green, 0; blue, 0 }  ,fill opacity=1 ] (229.19,265.97) -- (239.87,261.64) -- (240.72,267.4) -- cycle ;
\draw  [fill={rgb, 255:red, 0; green, 0; blue, 0 }  ,fill opacity=1 ] (116.41,98.59) -- (105.85,103.17) -- (104.86,97.43) -- cycle ;
\draw  [fill={rgb, 255:red, 0; green, 0; blue, 0 }  ,fill opacity=1 ] (217.03,98.47) -- (205.68,100.81) -- (206.01,95) -- cycle ;
\draw  [fill={rgb, 255:red, 0; green, 0; blue, 0 }  ,fill opacity=1 ] (85.3,272.21) -- (96.47,269.18) -- (96.54,275) -- cycle ;
\draw   (159.01,99.12) .. controls (157.04,100.15) and (155.17,101.14) .. (155.17,102.27) .. controls (155.18,103.41) and (157.06,104.38) .. (159.04,105.4) .. controls (161.01,106.41) and (162.89,107.39) .. (162.9,108.52) .. controls (162.9,109.66) and (161.03,110.65) .. (159.06,111.68) .. controls (157.1,112.71) and (155.22,113.7) .. (155.23,114.83) .. controls (155.23,115.97) and (157.11,116.94) .. (159.09,117.96) .. controls (161.07,118.97) and (162.95,119.95) .. (162.95,121.08) .. controls (162.96,122.22) and (161.09,123.21) .. (159.12,124.24) .. controls (157.15,125.27) and (155.28,126.26) .. (155.28,127.39) .. controls (155.29,128.53) and (157.17,129.5) .. (159.15,130.52) .. controls (161.12,131.53) and (163,132.51) .. (163.01,133.64) .. controls (163.01,134.78) and (161.14,135.77) .. (159.17,136.8) .. controls (157.21,137.83) and (155.33,138.82) .. (155.34,139.95) .. controls (155.34,141.09) and (157.22,142.06) .. (159.2,143.08) .. controls (161.18,144.09) and (163.06,145.07) .. (163.06,146.2) .. controls (163.07,147.34) and (161.2,148.33) .. (159.23,149.36) .. controls (157.26,150.39) and (155.39,151.38) .. (155.39,152.51) .. controls (155.4,153.65) and (157.28,154.62) .. (159.26,155.64) .. controls (161.23,156.66) and (163.11,157.63) .. (163.12,158.76) .. controls (163.12,159.9) and (161.25,160.89) .. (159.28,161.92) .. controls (157.32,162.95) and (155.44,163.94) .. (155.45,165.07) .. controls (155.45,166.21) and (157.33,167.18) .. (159.31,168.2) .. controls (160.14,168.63) and (160.96,169.05) .. (161.62,169.48) ;
\draw   (190.28,205.91) .. controls (188.34,206.94) and (186.5,207.93) .. (186.51,209.07) .. controls (186.51,210.2) and (188.36,211.17) .. (190.3,212.19) .. controls (192.24,213.21) and (194.09,214.18) .. (194.1,215.32) .. controls (194.1,216.45) and (192.26,217.44) .. (190.33,218.47) .. controls (188.4,219.5) and (186.56,220.49) .. (186.56,221.63) .. controls (186.57,222.76) and (188.42,223.73) .. (190.36,224.75) .. controls (192.3,225.77) and (194.15,226.74) .. (194.15,227.88) .. controls (194.16,229.01) and (192.32,230) .. (190.38,231.03) .. controls (188.45,232.06) and (186.61,233.05) .. (186.62,234.19) .. controls (186.62,235.32) and (188.47,236.29) .. (190.41,237.31) .. controls (192.35,238.33) and (194.2,239.3) .. (194.21,240.44) .. controls (194.21,241.57) and (192.37,242.56) .. (190.44,243.59) .. controls (188.51,244.62) and (186.67,245.61) .. (186.67,246.75) .. controls (186.68,247.88) and (188.53,248.86) .. (190.47,249.87) .. controls (192.41,250.89) and (194.26,251.86) .. (194.26,253) .. controls (194.27,254.13) and (192.43,255.12) .. (190.49,256.15) .. controls (188.56,257.18) and (186.72,258.17) .. (186.73,259.31) .. controls (186.73,260.44) and (188.58,261.42) .. (190.52,262.43) .. controls (192.46,263.45) and (194.31,264.42) .. (194.32,265.56) .. controls (194.32,266.69) and (192.48,267.68) .. (190.55,268.71) .. controls (189.76,269.14) and (188.98,269.55) .. (188.34,269.97) ;
\draw   (131.38,205.91) .. controls (129.78,206.94) and (128.25,207.92) .. (128.25,209.06) .. controls (128.26,210.2) and (129.79,211.17) .. (131.41,212.19) .. controls (133.02,213.21) and (134.56,214.18) .. (134.56,215.32) .. controls (134.57,216.45) and (133.04,217.44) .. (131.44,218.47) .. controls (129.83,219.5) and (128.3,220.48) .. (128.31,221.62) .. controls (128.31,222.76) and (129.85,223.73) .. (131.46,224.75) .. controls (133.08,225.77) and (134.61,226.74) .. (134.62,227.88) .. controls (134.62,229.01) and (133.1,230) .. (131.49,231.03) .. controls (129.89,232.06) and (128.36,233.04) .. (128.36,234.18) .. controls (128.37,235.32) and (129.9,236.29) .. (131.52,237.31) .. controls (133.13,238.33) and (134.67,239.3) .. (134.67,240.44) .. controls (134.68,241.57) and (133.15,242.56) .. (131.55,243.59) .. controls (129.94,244.62) and (128.41,245.61) .. (128.42,246.74) .. controls (128.42,247.88) and (129.96,248.85) .. (131.57,249.87) .. controls (133.19,250.89) and (134.72,251.86) .. (134.73,253) .. controls (134.73,254.13) and (133.21,255.12) .. (131.6,256.15) .. controls (130,257.18) and (128.47,258.17) .. (128.47,259.3) .. controls (128.48,260.44) and (130.01,261.41) .. (131.63,262.43) .. controls (133.24,263.45) and (134.78,264.42) .. (134.78,265.56) .. controls (134.79,266.69) and (133.26,267.68) .. (131.66,268.71) .. controls (130.05,269.74) and (128.52,270.73) .. (128.53,271.86) .. controls (128.53,272.07) and (128.58,272.27) .. (128.67,272.47) ;
\draw   (117.21,176.76) .. controls (117.21,172.73) and (120.47,169.47) .. (124.49,169.47) -- (194.5,169.47) .. controls (198.52,169.47) and (201.78,172.73) .. (201.78,176.76) -- (201.78,198.61) .. controls (201.78,202.63) and (198.52,205.9) .. (194.5,205.9) -- (124.49,205.9) .. controls (120.47,205.9) and (117.21,202.63) .. (117.21,198.61) -- cycle ;

\draw   (366.78,126.9) .. controls (310.28,82.9) and (675.51,74.62) .. (635.82,116.7) .. controls (596.14,158.78) and (598.83,190.65) .. (645.24,230.18) .. controls (691.65,269.71) and (314.99,286.93) .. (368.8,238.47) .. controls (422.61,190.02) and (423.28,170.89) .. (366.78,126.9) -- cycle ;
\draw  [fill={rgb, 255:red, 218; green, 24; blue, 24 }  ,fill opacity=1 ] (354,117.65) .. controls (354,113.96) and (357.16,110.96) .. (361.06,110.96) .. controls (364.96,110.96) and (368.12,113.96) .. (368.12,117.65) .. controls (368.12,121.35) and (364.96,124.35) .. (361.06,124.35) .. controls (357.16,124.35) and (354,121.35) .. (354,117.65) -- cycle ;
\draw  [fill={rgb, 255:red, 218; green, 24; blue, 24 }  ,fill opacity=1 ] (355.35,248.35) .. controls (355.35,244.66) and (358.51,241.66) .. (362.41,241.66) .. controls (366.31,241.66) and (369.47,244.66) .. (369.47,248.35) .. controls (369.47,252.05) and (366.31,255.05) .. (362.41,255.05) .. controls (358.51,255.05) and (355.35,252.05) .. (355.35,248.35) -- cycle ;
\draw  [fill={rgb, 255:red, 218; green, 24; blue, 24 }  ,fill opacity=1 ] (641.88,238.15) .. controls (641.88,234.46) and (645.04,231.46) .. (648.94,231.46) .. controls (652.84,231.46) and (656,234.46) .. (656,238.15) .. controls (656,241.85) and (652.84,244.85) .. (648.94,244.85) .. controls (645.04,244.85) and (641.88,241.85) .. (641.88,238.15) -- cycle ;
\draw  [fill={rgb, 255:red, 218; green, 24; blue, 24 }  ,fill opacity=1 ] (628.76,110) .. controls (628.76,106.3) and (631.92,103.31) .. (635.82,103.31) .. controls (639.72,103.31) and (642.88,106.3) .. (642.88,110) .. controls (642.88,113.7) and (639.72,116.7) .. (635.82,116.7) .. controls (631.92,116.7) and (628.76,113.7) .. (628.76,110) -- cycle ;
\draw   (481.8,169.61) .. controls (481.8,166.8) and (484.08,164.51) .. (486.9,164.51) -- (523.78,164.51) .. controls (526.59,164.51) and (528.88,166.8) .. (528.88,169.61) -- (528.88,184.92) .. controls (528.88,187.73) and (526.59,190.02) .. (523.78,190.02) -- (486.9,190.02) .. controls (484.08,190.02) and (481.8,187.73) .. (481.8,184.92) -- cycle ;
\draw   (390.32,129.45) .. controls (363.42,88.01) and (653.31,86.73) .. (618.33,126.9) .. controls (583.36,167.06) and (598.83,203.41) .. (631.79,228.27) .. controls (664.74,253.14) and (351.98,281.83) .. (393.68,231.46) .. controls (435.39,181.09) and (417.22,170.89) .. (390.32,129.45) -- cycle ;
\draw   (510.04,158.14) -- (535.6,158.14) -- (535.6,196.39) ;
\draw   (475.74,196.39) -- (475.74,158.58) -- (501.3,158.14) ;
\draw   (475.74,169.1) -- (475.74,196.46) -- (485.16,196.63) ;
\draw   (535.6,169.73) -- (535.6,196.39) -- (526.86,196.55) ;
\draw    (485.16,196.63) -- (484.49,259.83) ;
\draw    (492.56,197.03) -- (492.56,258.24) ;
\draw    (501.3,97.57) -- (501.3,158.14) ;
\draw    (509.37,97.57) -- (509.37,158.14) ;
\draw    (526.86,196.39) -- (526.86,256.32) ;
\draw    (517.44,197.03) -- (517.44,257.6) ;
\draw    (492.56,197.03) -- (517.44,197.03) ;
\draw [color={rgb, 255:red, 255; green, 255; blue, 255 }  ,draw opacity=1 ]   (501.3,97.57) -- (509.37,97.57) ;
\draw [color={rgb, 255:red, 255; green, 255; blue, 255 }  ,draw opacity=1 ][line width=2.25]    (517.44,257.6) -- (526.86,256.96) ;
\draw [color={rgb, 255:red, 255; green, 255; blue, 255 }  ,draw opacity=1 ][line width=2.25]    (484.49,259.83) -- (492.56,258.24) ;
\draw  [fill={rgb, 255:red, 0; green, 0; blue, 0 }  ,fill opacity=1 ] (468.2,99.52) -- (457.58,104.26) -- (456.52,98.44) -- cycle ;
\draw  [fill={rgb, 255:red, 0; green, 0; blue, 0 }  ,fill opacity=1 ] (554.37,98.81) -- (542.9,101.18) -- (543.23,95.29) -- cycle ;
\draw  [fill={rgb, 255:red, 0; green, 0; blue, 0 }  ,fill opacity=1 ] (465.51,90.6) -- (454.89,95.34) -- (453.83,89.52) -- cycle ;
\draw  [fill={rgb, 255:red, 0; green, 0; blue, 0 }  ,fill opacity=1 ] (554.37,90.53) -- (542.9,92.89) -- (543.23,87) -- cycle ;
\draw  [fill={rgb, 255:red, 0; green, 0; blue, 0 }  ,fill opacity=1 ] (501.34,133.22) -- (498.15,122.51) -- (504.38,122.47) -- cycle ;
\draw  [fill={rgb, 255:red, 0; green, 0; blue, 0 }  ,fill opacity=1 ] (485.2,231.4) -- (482,220.7) -- (488.23,220.65) -- cycle ;
\draw  [fill={rgb, 255:red, 0; green, 0; blue, 0 }  ,fill opacity=1 ] (492.66,220.36) -- (495.56,231.14) -- (489.34,231.03) -- cycle ;
\draw  [fill={rgb, 255:red, 0; green, 0; blue, 0 }  ,fill opacity=1 ] (509.48,122.49) -- (512.38,133.27) -- (506.15,133.16) -- cycle ;
\draw  [fill={rgb, 255:red, 0; green, 0; blue, 0 }  ,fill opacity=1 ] (526.96,220.99) -- (529.87,231.78) -- (523.64,231.67) -- cycle ;
\draw  [fill={rgb, 255:red, 0; green, 0; blue, 0 }  ,fill opacity=1 ] (517.5,233.96) -- (514.27,223.25) -- (520.5,223.2) -- cycle ;
\draw  [fill={rgb, 255:red, 0; green, 0; blue, 0 }  ,fill opacity=1 ] (511.61,189.75) -- (500.3,192.71) -- (500.29,186.81) -- cycle ;

\draw   (297,181) -- (314.5,170) -- (314.5,175.5) -- (349.5,175.5) -- (349.5,170) -- (367,181) -- (349.5,192) -- (349.5,186.5) -- (314.5,186.5) -- (314.5,192) -- cycle ;

\end{tikzpicture}
}
 
 \caption[]
{Diagram for the $\pi \pi$ scattering amplitude including an internal 2 index quark loop that should mimick a four quark state intermediate state. } \label{FigD}
\end{figure}

One of the fascinating aspects of the two-index antisymmetric large $N$ limit is that it offers a physical way to understand why unitarity is achieved faster in $N$ than in the 't Hooft limit. This is best explained by examining the large $N$ dynamics of the diagram in Fig.~\ref{FigD}.
The diagram accounts for four pion insertions, six gauge couplings and five closed loops. With the rules detailed above the large $N$ dynamics yields a contribution to the amplitude that goes like $1/N^2$ which is of the same order as the diagram without internal (two-index) quark loops. As expected \cite{Kiritsis:1989ge,Sannino:2007yp} two-index quarks behave, at leading order in $1/N$, as gluons in scattering amplitudes. As observed in \cite{Sannino:2007yp} it is natural to interpret the contribution to the $\pi\pi$ amplitude stemming from 
the diagram in  Fig.~\ref{FigD} as an intermediate state made by four quarks.  Thus in the CR large $N$ limits, differently from the 't Hooft one, {\it exotic} four (and multi) quark states appear already at the leading order alongside the two quark resonances. This is very much what is needed to unitarize the physical $\pi \pi$ amplitude  
for $N=3$ \cite{Harada:1995dc,Harada:1996wr}. In fact, already some time ago, in \cite{Sannino:1995ik} we showed that in the 't Hooft  limit the leading quark-antiquark resonances are unable to unitarize $\pi \pi$ scattering. About ten months later \cite{Harada:1995dc} we further  argued that the sigma state, made by more than two quarks (a picture consistent with Jaffe's model \cite{Jaffe:1976ig}), was crucial to unitarize the isospin and angular momentum zero amplitude. Recently the 't Hooft extrapolation for pion pion scattering has been discussed via lattice simulations \cite{Baeza-Ballesteros:2022azb,Baeza-Ballesteros:2024ogp,Hernandez:2020tbc,Hernandez:2019qed} and earlier on via different non-perturbative unitarization methods in \cite{Truong:1988zp,Truong:1988zp,Truong:1988zp,Nieves:2011gb} and more recently via sum rules in \cite{Cid-Mora:2022kgu}.  Given the agreement with phenomenological evidence one concludes that: 

\vskip .1cm
\noindent 
{\it The two-index antisymmetric large $N$ limit of ordinary QCD with light flavours is a more realistic and converging counting scheme for low energy meson scattering amplitudes than the 't Hooft extrapolation.} 
\vskip .1cm
{\it In the CR limit tetraquark states appear at leading order in the meson-meson amplitude and help
unitarize the effective theory as observed in $\pi \pi$ scattering.}
\vskip .1cm

Of course, this does not mean that one should discount the 't Hooft extrapolation. The latter best addresses the complementary gluon-rich dynamics including the fact that   meson states are dominated by quark-antiquark configurations, the understanding of the OZI rule and the fact that baryons can be seen to emerge in an elegant way as solitons in the model. 

\begin{figure}[htbp]
\centering 

\tikzset{every picture/.style={line width=0.75pt}} 
\scalebox{0.70}{

\begin{tikzpicture}[x=0.75pt,y=0.75pt,yscale=-1,xscale=1]

\draw  [fill={rgb, 255:red, 0; green, 0; blue, 0 }  ,fill opacity=1 ] (90.59,121.28) -- (75.98,128.73) -- (74.05,121) -- cycle ;
\draw  [fill={rgb, 255:red, 0; green, 0; blue, 0 }  ,fill opacity=1 ] (82.55,241.55) -- (98.41,237.07) -- (98.7,245.01) -- cycle ;
\draw   (142.61,145.3) .. controls (143.7,147.34) and (144.74,149.27) .. (145.96,149.27) .. controls (147.17,149.27) and (148.22,147.34) .. (149.31,145.3) .. controls (150.4,143.27) and (151.45,141.33) .. (152.66,141.33) .. controls (153.88,141.33) and (154.92,143.27) .. (156.01,145.3) .. controls (157.1,147.34) and (158.15,149.27) .. (159.36,149.27) .. controls (160.57,149.27) and (161.62,147.34) .. (162.71,145.3) .. controls (163.81,143.27) and (164.85,141.33) .. (166.07,141.33) .. controls (167.28,141.33) and (168.32,143.27) .. (169.42,145.3) .. controls (170.51,147.34) and (171.55,149.27) .. (172.76,149.27) .. controls (173.98,149.27) and (175.02,147.34) .. (176.12,145.3) .. controls (177.21,143.27) and (178.26,141.33) .. (179.47,141.33) .. controls (180.68,141.33) and (181.73,143.27) .. (182.82,145.3) .. controls (183.91,147.34) and (184.95,149.27) .. (186.17,149.27) .. controls (187.38,149.27) and (188.43,147.34) .. (189.52,145.3) .. controls (190.61,143.27) and (191.66,141.33) .. (192.87,141.33) .. controls (194.09,141.33) and (195.13,143.27) .. (196.22,145.3) .. controls (197.31,147.34) and (198.36,149.27) .. (199.57,149.27) .. controls (200.78,149.27) and (201.83,147.34) .. (202.92,145.3) .. controls (204.02,143.27) and (205.06,141.33) .. (206.28,141.33) .. controls (207.49,141.33) and (208.53,143.27) .. (209.63,145.3) .. controls (210.72,147.34) and (211.76,149.27) .. (212.97,149.27) .. controls (214.19,149.27) and (215.23,147.34) .. (216.33,145.3) .. controls (216.79,144.45) and (217.24,143.61) .. (217.7,142.92) ;
\draw   (141.54,218.5) .. controls (142.63,220.54) and (143.68,222.47) .. (144.89,222.47) .. controls (146.1,222.47) and (147.15,220.54) .. (148.24,218.5) .. controls (149.34,216.47) and (150.38,214.53) .. (151.6,214.53) .. controls (152.81,214.53) and (153.85,216.47) .. (154.95,218.5) .. controls (156.04,220.54) and (157.08,222.47) .. (158.29,222.47) .. controls (159.51,222.47) and (160.55,220.54) .. (161.65,218.5) .. controls (162.74,216.47) and (163.79,214.53) .. (165,214.53) .. controls (166.21,214.53) and (167.26,216.47) .. (168.35,218.5) .. controls (169.44,220.54) and (170.48,222.47) .. (171.7,222.47) .. controls (172.91,222.47) and (173.96,220.54) .. (175.05,218.5) .. controls (176.14,216.47) and (177.19,214.53) .. (178.4,214.53) .. controls (179.62,214.53) and (180.66,216.47) .. (181.75,218.5) .. controls (182.84,220.54) and (183.89,222.47) .. (185.1,222.47) .. controls (186.31,222.47) and (187.36,220.54) .. (188.45,218.5) .. controls (189.55,216.47) and (190.59,214.53) .. (191.81,214.53) .. controls (193.02,214.53) and (194.06,216.47) .. (195.15,218.5) .. controls (196.25,220.54) and (197.29,222.47) .. (198.5,222.47) .. controls (199.72,222.47) and (200.76,220.54) .. (201.86,218.5) .. controls (202.95,216.47) and (204,214.53) .. (205.21,214.53) .. controls (206.42,214.53) and (207.47,216.47) .. (208.56,218.5) .. controls (209.65,220.54) and (210.69,222.47) .. (211.91,222.47) .. controls (213.12,222.47) and (214.17,220.54) .. (215.26,218.5) .. controls (215.72,217.65) and (216.17,216.81) .. (216.63,216.12) ;
\draw   (28.46,181.27) .. controls (28.48,148.13) and (56.29,121.27) .. (90.59,121.28) .. controls (124.89,121.29) and (152.69,148.17) .. (152.68,181.31) .. controls (152.66,214.45) and (124.85,241.31) .. (90.55,241.29) .. controls (56.25,241.28) and (28.45,214.41) .. (28.46,181.27) -- cycle ;
\draw   (220.7,146.14) .. controls (218.6,147.2) and (216.6,148.21) .. (216.6,149.38) .. controls (216.6,150.55) and (218.6,151.56) .. (220.7,152.62) .. controls (222.81,153.68) and (224.81,154.69) .. (224.81,155.86) .. controls (224.81,157.03) and (222.81,158.04) .. (220.7,159.09) .. controls (218.6,160.15) and (216.6,161.16) .. (216.6,162.33) .. controls (216.6,163.5) and (218.6,164.51) .. (220.7,165.57) .. controls (222.81,166.63) and (224.81,167.64) .. (224.81,168.81) .. controls (224.81,169.98) and (222.81,170.99) .. (220.7,172.04) .. controls (218.6,173.1) and (216.6,174.11) .. (216.6,175.28) .. controls (216.6,176.45) and (218.6,177.46) .. (220.7,178.52) .. controls (222.81,179.58) and (224.81,180.59) .. (224.81,181.76) .. controls (224.81,182.93) and (222.81,183.94) .. (220.7,184.99) .. controls (218.6,186.05) and (216.6,187.06) .. (216.6,188.23) .. controls (216.6,189.4) and (218.6,190.41) .. (220.7,191.47) .. controls (222.81,192.53) and (224.81,193.54) .. (224.81,194.71) .. controls (224.81,195.88) and (222.81,196.89) .. (220.7,197.94) .. controls (218.6,199) and (216.6,200.01) .. (216.6,201.18) .. controls (216.6,202.35) and (218.6,203.36) .. (220.7,204.42) .. controls (222.81,205.48) and (224.81,206.49) .. (224.81,207.66) .. controls (224.81,208.83) and (222.81,209.84) .. (220.7,210.89) .. controls (218.6,211.95) and (216.6,212.96) .. (216.6,214.13) .. controls (216.6,215.3) and (218.6,216.31) .. (220.7,217.37) .. controls (221.59,217.81) and (222.46,218.25) .. (223.17,218.69) ;
\draw  [fill={rgb, 255:red, 218; green, 24; blue, 24 }  ,fill opacity=1 ] (21,182) .. controls (21,178.24) and (24.34,175.2) .. (28.46,175.2) .. controls (32.58,175.2) and (35.92,178.24) .. (35.92,182) .. controls (35.92,185.75) and (32.58,188.8) .. (28.46,188.8) .. controls (24.34,188.8) and (21,185.75) .. (21,182) -- cycle ;
\draw  [fill={rgb, 255:red, 29; green, 33; blue, 189 }  ,fill opacity=1 ] (210.95,149.38) .. controls (210.95,145.62) and (214.29,142.58) .. (218.41,142.58) .. controls (222.53,142.58) and (225.87,145.62) .. (225.87,149.38) .. controls (225.87,153.13) and (222.53,156.18) .. (218.41,156.18) .. controls (214.29,156.18) and (210.95,153.13) .. (210.95,149.38) -- cycle ;
\draw  [fill={rgb, 255:red, 29; green, 33; blue, 189 }  ,fill opacity=1 ] (213.08,217.42) .. controls (213.08,213.67) and (216.42,210.62) .. (220.55,210.62) .. controls (224.67,210.62) and (228.01,213.67) .. (228.01,217.42) .. controls (228.01,221.18) and (224.67,224.22) .. (220.55,224.22) .. controls (216.42,224.22) and (213.08,221.18) .. (213.08,217.42) -- cycle ;

\draw   (376.77,129.2) .. controls (430.64,79.19) and (658.98,84.25) .. (622.7,120.43) .. controls (586.43,156.6) and (588.89,184.01) .. (631.31,217.99) .. controls (673.73,251.97) and (434.33,262.47) .. (378.61,225.12) .. controls (322.89,187.77) and (322.9,179.21) .. (376.77,129.2) -- cycle ;
\draw  [fill={rgb, 255:red, 218; green, 24; blue, 24 }  ,fill opacity=1 ] (330.02,182.08) .. controls (330.02,178.9) and (332.91,176.32) .. (336.47,176.32) .. controls (340.04,176.32) and (342.93,178.9) .. (342.93,182.08) .. controls (342.93,185.26) and (340.04,187.83) .. (336.47,187.83) .. controls (332.91,187.83) and (330.02,185.26) .. (330.02,182.08) -- cycle ;
\draw   (389.09,149.18) .. controls (389.09,139.45) and (396.98,131.55) .. (406.72,131.55) -- (464.71,131.55) .. controls (474.45,131.55) and (482.34,139.45) .. (482.34,149.18) -- (482.34,202.07) .. controls (482.34,211.81) and (474.45,219.7) .. (464.71,219.7) -- (406.72,219.7) .. controls (396.98,219.7) and (389.09,211.81) .. (389.09,202.07) -- cycle ;
\draw   (499.89,125.48) .. controls (499.89,116.3) and (507.33,108.86) .. (516.5,108.86) -- (566.36,108.86) .. controls (575.54,108.86) and (582.98,116.3) .. (582.98,125.48) -- (582.98,212.68) .. controls (582.98,221.86) and (575.54,229.3) .. (566.36,229.3) -- (516.5,229.3) .. controls (507.33,229.3) and (499.89,221.86) .. (499.89,212.68) -- cycle ;
\draw  [fill={rgb, 255:red, 0; green, 0; blue, 0 }  ,fill opacity=1 ] (442.65,102.4) -- (430.01,108.7) -- (428.34,102.16) -- cycle ;
\draw  [fill={rgb, 255:red, 0; green, 0; blue, 0 }  ,fill opacity=1 ] (440.07,131.67) -- (426.32,135.39) -- (426.11,128.66) -- cycle ;
\draw  [fill={rgb, 255:red, 0; green, 0; blue, 0 }  ,fill opacity=1 ] (482.03,175.92) -- (478.57,162.81) -- (485.68,162.84) -- cycle ;
\draw  [fill={rgb, 255:red, 0; green, 0; blue, 0 }  ,fill opacity=1 ] (500.5,176.79) -- (497.03,163.68) -- (504.15,163.71) -- cycle ;
\draw  [fill={rgb, 255:red, 0; green, 0; blue, 0 }  ,fill opacity=1 ] (551.79,92.78) -- (537.83,95.73) -- (538.04,89) -- cycle ;
\draw  [fill={rgb, 255:red, 0; green, 0; blue, 0 }  ,fill opacity=1 ] (529.63,249.01) -- (543.35,245.22) -- (543.6,251.94) -- cycle ;
\draw  [fill={rgb, 255:red, 0; green, 0; blue, 0 }  ,fill opacity=1 ] (430.01,241.59) -- (444.26,240.46) -- (443.08,247.1) -- cycle ;
\draw  [fill={rgb, 255:red, 29; green, 33; blue, 189 }  ,fill opacity=1 ] (621.02,113.25) .. controls (621.02,109.61) and (624.15,106.66) .. (628.01,106.66) .. controls (631.88,106.66) and (635.01,109.61) .. (635.01,113.25) .. controls (635.01,116.89) and (631.88,119.85) .. (628.01,119.85) .. controls (624.15,119.85) and (621.02,116.89) .. (621.02,113.25) -- cycle ;
\draw  [fill={rgb, 255:red, 29; green, 33; blue, 189 }  ,fill opacity=1 ] (628.02,228.25) .. controls (628.02,224.61) and (631.15,221.66) .. (635.01,221.66) .. controls (638.88,221.66) and (642.01,224.61) .. (642.01,228.25) .. controls (642.01,231.89) and (638.88,234.85) .. (635.01,234.85) .. controls (631.15,234.85) and (628.02,231.89) .. (628.02,228.25) -- cycle ;

\draw   (246,183) -- (263.5,172) -- (263.5,177.5) -- (298.5,177.5) -- (298.5,172) -- (316,183) -- (298.5,194) -- (298.5,188.5) -- (263.5,188.5) -- (263.5,194) -- cycle ;

\end{tikzpicture}
}
\caption[]
{Diagram for a meson decay (red dot) into two glueballs (blue dots).} \label{FigE}
\end{figure}
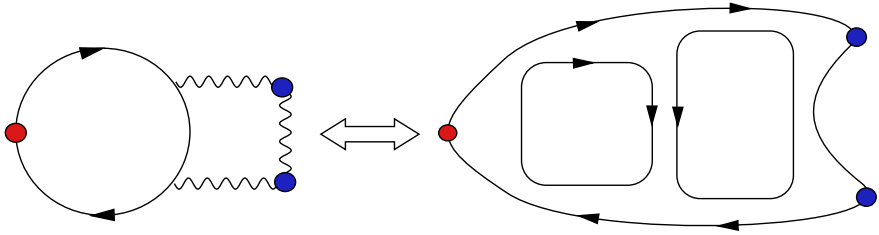

For the large $N$ rules for the meson 
and glueball masses and decays in the two-index theories one has that: both the
 glueball and meson are ${\cal O}(N^0)$ and that the decay of a meson into two glueballs 
scale as $1/N$ as shown in Fig.\ref{FigE}. Recall that the
glueball insertion (blue dot) scales as $1/N$ and that two QCD interaction vertices are involved.

\vskip .2cm
It is now clear that Coleman analysis is consistent with 't Hooft large $N$ limit but that the physics of tetraquarks is best captured via the CR limit, in other words it approaches faster in $N$ the phenomenological result. 
 
Each extrapolation emphasizes complementary aspects of QCD and the not unique nature of the large $N$ extrapolation means that they are not a replacement for the whole theory.

\section{Epilogue}
\label{summary}

I reviewed salient properties of large $N$ dynamics of two-index $SU(N)$ gauge theories  that help us access corners of strongly interacting dynamics missed by the 't Hooft extrapolation. Such a revival is  partially triggered by recent precise lattice results about the spectrum of these theories \cite{Ziegler:2021nbl,DellaMorte:2023ylq,DellaMorte:2023sdz,Jaeger:2022ypq}. The distinct group-theoretical ways one can depart from  QCD, of either vector  or chiral nature, for arbitrary number of flavors were introduced. These  are the 't Hooft \cite{tHooft:1973alw}, CR \cite{Corrigan:1979xf} and our chiral extension of \cite{Sannino:2005sk}.   Furthermore, one flavor QCD theory  in the CR limit is planar equivalent to super Yang-Mills \cite{Armoni:2003fb,Armoni:2003gp} with the associated finite $N$ effective lagrangian theory was constructed in \cite{Sannino:2003xe}. Because the  finite $N$ spectrum of two-index antisymmetric, and possibly in the future also of the two-index symmetric, theories is being studied via lattice simulations I re-examined the predictions stemming from the effective approach  \cite{Sannino:2003xe}. At finite number of colors the pseudoscalar to scalar ratio departs from unity and the leading $1/N$ corrections were estimated in \cite{Sannino:2003xe}. Here I highlighted the dependence of the ratio in terms of the axial and trace anomalies and further motivated an improved  finite $N$ version of the results expressed as: 
 \begin{eqnarray}
\frac{M_{\eta^{\prime}}}{M_{\sigma}}\Big|_{\rm improved} = 
 \frac{1 \pm {2}/{N}}{ 
{\beta_{O\pm}}/{\beta_{SYM}} + \frac{4}{9} b} -  \frac{m}{\alpha \lambda \Lambda}+
{O}(m^2)\leq   \frac{1 \pm {2}/{N}}{ 1 \mp \frac{4}{9N}  }-  \frac{m}{\alpha \lambda \Lambda}+ 
{O}(m^2)\ .
\end{eqnarray}
We retained the leading correction in the fermion mass which stems from a protected soft supersymmetry breaking operator. In fact, in absence of an experimental evidence of supersymmetry, simulations of this ratio as function of fermion mass constitute a direct way to measure soft supersymmetry breaking.  I   also commented on a string theory inspired estimate of the pseudoscalar to scalar mass ratio \cite{Armoni:2005qr}. I then moved to determine the size of the $1/N^2$ corrections by a direct comparison with lattice results \cite{DellaMorte:2023sdz} and shown that they are remarkably under control already for $N=3$, in the original $1/N$ estimate of \cite{Sannino:2003xe} which was supposed to hold only in the vicinity of $N\rightarrow \infty$. Other quantities, computed at finite $N$ in \cite{Sannino:2003xe}, deserve attention such as the gluon condensate, the vacuum energy and its $\theta$-angle dependence as they reveal sought after aspects of strongly coupled quantum field theories.

When adding flavors the phase
diagram as a function of the number of flavors and
 colors has been provided in \cite{Sannino:2004qp}. The  complete phase diagram for fermions in arbitrary representations has been unveiled in \cite{Dietrich:2006cm}. The study of theories with fermions in a higher dimensional
 representation of the gauge group and
 the knowledge of the associated conformal window led to
the construction of minimal models of technicolor
\cite{Sannino:2004qp,hep-ph/0406200, Dietrich:2005jn}. A recent summary can be found in \cite{Cacciapaglia:2020kgq}.

An exciting are of research  is  meson-meson scattering for its theoretical aspects and phenomenological impact, such as the study of exotic\footnote{Exotic is referred to the 't Hooft extrapolation since in the CR limit these states are leading in $N$.} states like tetraquarks. The subject has a long history with the importance of tetraquarks for two and three flavors QCD dating back to the pioneering work of Jaffe \cite{Jaffe:1976ig}  and the problems associated to the 't Hooft extrapolation for pion pion scattering noted in  \cite{Sannino:1995ik} and then resolved by the inclusion of tetraquarks in \cite{Harada:1995dc,Harada:1996wr,Black:1998wt}. Recently, tetraquark states have also been  discovered by the LHCb collaboration \cite{LHCb:2020bls,LHCb:2020pxc,LHCb:2022lzp,LHCb:2022sfr} with the $f_0(500)$ further investigated by BESIII \cite{BESIII:2024lnh}.  At the same time  lattice simulations of meson-meson scattering \cite{Hernandez:2019qed,Hernandez:2020tbc,Baeza-Ballesteros:2022azb,Baeza-Ballesteros:2024ogp} apt at understanding the limitations of the 't Hooft large $N$ limit as well as the emergence of tetraquark states have appeared. 

To homage Weinberg's brilliant work, I reviewed his short paper on tetraquark states and large $N$ meson-meson dynamics \cite{Weinberg:2013cfa},  and then employed  the CR limit to address some of the questions posed in his work.  
 
I can now summarize the main features recommending   the two-index antisymmetric large $N$ extension of multiflavor  QCD over the 't Hooft extrapolation:   

\begin{itemize}

\item[i)] {It provides a better description of the large $N$ properties meson-meson scattering. For example, the associated scattering amplitudes unitarize more rapidly when increasing $N$;}

\item [ii)] {Tetraquark states appear at leading order in the meson-meson amplitude and help unitarize the effective theory at intermediate energies between the Goldstone realm and the higher spin resonances as observed in $\pi \pi$ scattering. } 


\end{itemize}
It is therefore highly interesting to perform lattice investigations similar to the ones performed in   \cite{Hernandez:2019qed,Hernandez:2020tbc,Baeza-Ballesteros:2022azb,Baeza-Ballesteros:2024ogp}  for $N=5$ or higher number of colors featuring two Dirac light flavors in the two-index antisymmetric representation. In these simulations the  tetraquark states should remain in the spectrum at lower energy and become narrower. In other words one can use the CR limit to single out these states. 

 \vskip .4cm
 Concluding, Weinberg was more than a scientist; he was a visionary who brilliantly bridged theory and experiments, guiding us towards a unified understanding of nature's fundamental forces. His work remains a cornerstone of modern physics, reshaping our comprehension of the laws of Nature. Weinberg's journey may have ended, but his work continues to guide us toward new discoveries and understanding, a beacon of light for future generations to follow.
 
\Acknowledgements

\noindent  
 
The work of F.S. is
partially supported by the Carlsberg Foundation, semper ardens grant CF22-0922.

\clearpage
\bibliographystyle{JHEP}      
\bibliography{bibtexref.bib}
 \end{document}